\documentclass[twocolumn]{aastex631}
\usepackage{newtxtext,newtxmath}
\shorttitle{Coupled evolution of Earth's proto-atmosphere and magma ocean}
\shortauthors{Yoshida et al.}
\graphicspath{{./}{figures/}}

\begin{document}

\title{Coupled evolution of Earth’s H$_2$-rich proto-atmosphere and oxidized magma ocean: Efficient H$_2$O production through dissolved H$_2$ transport and atmospheric buffering of the magma ocean redox state}

\correspondingauthor{Tatsuya Yoshida}
\email{tatsuya@elsi.jp}

\author[0009-0003-0399-1305]{Tatsuya Yoshida}
\affiliation{Earth-Life Science Institute, Institute of Science Tokyo, Tokyo, Japan}
\affiliation{Department of Earth and Space Sciences, University of Washington, Seattle, WA, USA}
\affiliation{Graduate School of Science, Tohoku University, Miyagi, Japan.}

\author{Kirara Arima}
\affiliation{Faculty of Science, Hokkaido University, Hokkaido, Japan}

\author[0000-0001-5685-9736]{Naoki Terada}
\affiliation{Graduate School of Science, Tohoku University, Miyagi, Japan.}

\author[0000-0003-4789-4737]{Takeshi Kuroda}
\affiliation{National Institute of Information and Communications Technology, Tokyo, Japan}
\affiliation{Graduate School of Science, Tohoku University, Miyagi, Japan.}
\affiliation{Division for the Establishment of Frontier Sciences of Organization for Advanced Studies, Tohoku University, Miyagi, Japan}

\author[0000-0002-6757-8064]{Kiyoshi Kuramoto}
\affiliation{Faculty of Science, Hokkaido University, Hokkaido, Japan}



\begin{abstract}
	Giant impacts during Earth’s final accretion likely generated global magma oceans, where the disproportionation of Fe$^{2+}$ at lower-mantle pressures and segregation of the resulting metallic iron into the core could have oxidized the magma ocean. Earth may nevertheless have had an H$_2$-rich proto-atmosphere in chemical disequilibrium with the magma ocean, but the chemical coupling between this atmosphere and an oxidized magma ocean remains poorly constrained. Here we develop a coupled atmosphere--interior evolution model that includes diffusive transport in the boundary layer between the magma ocean and atmosphere, redox reactions in the melt, partitioning of oxide components and H$_2$O among coexisting phases, and planetary cooling through a proto-atmosphere, with metallic iron produced by Fe$^{2+}$ disproportionation removed into the core as it forms. Atmospheric H$_2$ transport into the magma ocean and subsequent H$_2$O production become more efficient with increasingly frequent boundary layer renewal driven by buoyant overturn of dense melt cooled at the surface, allowing the atmosphere and magma ocean to chemically equilibrate within $\sim10^4$ yr when the magma ocean is sufficiently deep and hot to extend into the lower mantle. Redox exchange with the atmosphere buffers the solidification-driven redox evolution of the magma ocean, preventing extreme oxidation or reduction of the coupled system. These results suggest that the coupled evolution could have contributed to early Earth’s water inventory while maintaining a reducing H$_2$-rich atmosphere.
\end{abstract}



\section{Introduction} \label{sec:intro}
	 Earth is thought to have experienced a globally molten state, due to giant impacts during the late stage of accretion \citep[e.g.,][]{Canup2004a,Solomatov2007,Zahnle2007,Elkins2012}. Although the magma ocean is expected to have initially been highly reducing, with a redox state buffered by coexistence with metallic iron \citep{Frost2008,Hirschmann2012}, the formation of a deep magma ocean would have allowed ferrous iron (Fe$^{2+}$ or FeO) to undergo disproportionation at lower-mantle pressures, producing both metallic iron (Fe) and ferric iron (Fe$^{3+}$ or FeO$_{1.5}$), and the subsequent segregation of the resulting metallic iron into the core and convective homogenization of Fe$^{3+}$ throughout the magma ocean would have oxidized the magma ocean to a redox state comparable to that of the present-day mantle \citep{Hirschmann2012,Zhang2017,Armstrong2019,Deng2020,Hirschmann2022,Kuwahara2023,Zhang2024}.
	 
	 At the same time, Earth is expected to have acquired and retained a proto-atmosphere through degassing from accreting materials and the interior, together with the retention of captured solar nebular gas \citep[e.g.,][]{Hayashi1979,Matsui1986a}. Like the magma ocean, this proto-atmosphere could also initially have been highly reducing and H$_2$-rich, with its redox state buffered by coexistence with metallic iron during core formation \citep[e.g.,][]{Urey1952,Kuramoto1996,Schaefer2010,Hirschmann2012}. However, the oxidation of the magma ocean through Fe$^{2+}$ disproportionation and segregation of metallic iron into the core would have affected the composition of the proto-atmosphere via chemical interaction at the interface. Previous studies investigating the thermochemical equilibrium between a magma ocean and an overlying proto-atmosphere have indicated that even highly reduced atmospheres can become enriched in oxidized species such as H$_2$O, CO, and CO$_2$ \citep{Hirschmann2012,Pahlevan2019,Deng2020,Sossi2020,Katyal2020,Bower2022,Gaillard2022,Maurice2024,Nicholls2024,Krissansen2024,Zhang2024,Horn2025,Miozzi2025}. An important implication of these results is that H$_2$O can be generated through redox reactions between an oxidized magma ocean and a reduced proto-atmosphere. Moreover, Fe$^{3+}$ can become increasingly enriched in the magma ocean as the crystallization proceeds due to its incompatibility with mantle minerals \citep{Schaefer2024}, further elevating the oxygen fugacity of the magma ocean and leading to the formation of a strongly oxidized atmosphere. 
	 
	 However, how a magma ocean oxidized by the disproportionation of Fe$^{2+}$ and subsequent segregation of the resulting Fe into the core affects an H$_2$-rich reduced proto-atmosphere remains incompletely understood. One remaining uncertainty is the chemical feedback from the atmosphere on the redox evolution of the solidifying magma ocean. Previous magma ocean--atmosphere models for Earth have often assumed that the mantle and magma ocean act as effectively infinite oxygen reservoirs, thereby neglecting the oxygen mass balance associated with redox reactions between the atmosphere and magma ocean \citep[][]{Hirschmann2012,Pahlevan2019,Katyal2020,Bower2022,Maurice2024,Nicholls2024}, as pointed out by \citet{Krissansen2024}, and derive the atmospheric composition as a dependent variable determined by the redox state of the magma ocean. Although this assumption is reasonable when the oxygen mass in the planetary interior greatly exceeds that in the atmosphere, redox reactions with the atmosphere can consume oxidizing capacity in the magma ocean and suppress the solidification-driven enhancement of the Fe$^{3+}$ concentration once the atmospheric mass is non-negligible or as the magma ocean solidification proceeds. This modification of the magma ocean redox state can also alter the atmospheric composition and surface environment just after the solidification. Because \citet{Krissansen2024} investigated the co-evolution of an H$_2$-rich primary atmosphere and a magma ocean without focusing on magma ocean oxidation by Fe$^{2+}$ disproportionation, whereas \citet{Schaefer2024} estimated solidification-driven Fe$^{3+}$ enrichment without explicitly treating chemical modification by the atmosphere, it remains unclear how an H$_2$-rich atmosphere modifies the redox evolution of a solidifying magma ocean that is oxidized by Fe$^{2+}$ disproportionation and subsequent Fe$^{3+}$ enrichment.
	
	Moreover, the rate and efficiency of the chemical interaction between the magma ocean and atmosphere remain poorly constrained. In particular, the overall rate of H$_2$O production should depend on the transport of atmospheric H$_2$ across the boundary layer between the atmosphere and magma ocean \citep{Olson2018,Olson2019}. This redox interaction would have proceeded in parallel with other key processes affecting atmospheric evolution, including planetary cooling via outgoing thermal radiation and hydrogen escape to space. Therefore, understanding the timescale of the chemical coupling is essential for understanding the co-evolution of the proto-atmosphere and magma ocean during this stage.  

	To address these unresolved issues, we develop a coupled atmosphere-interior evolution model that incorporates the diffusion of atmospheric H$_2$ into the magma ocean, redox reactions within the melt, and partitioning of oxide components and H$_2$O among coexisting phases, together with planetary cooling governed by radiative transfer through the proto-atmosphere. Using this model, we investigate how a magma ocean oxidized by Fe$^{2+}$ disproportionation, with continuous removal of the resulting Fe into the core as it forms, affects an overlying H$_2$-rich proto-atmosphere during the post-core-formation stage. Specifically, we quantify the atmospheric influence on the redox state of the solidifying magma ocean while accounting for material exchange between the magma ocean and solid mantle, including uncertainty in its efficiency, and constrain the timescale for chemical interaction between the atmosphere and magma ocean. The remainder of this paper is organized as follows. In Section 2, we describe the coupled evolution model. Section 3 provides the calculation results, while Section 4 discusses model uncertainties, implications for present-day Earth and Mars, and the subsequent atmospheric evolution after magma ocean solidification.  

\begin{figure*}[htbp]
	\centering
	\includegraphics[width=\textwidth]{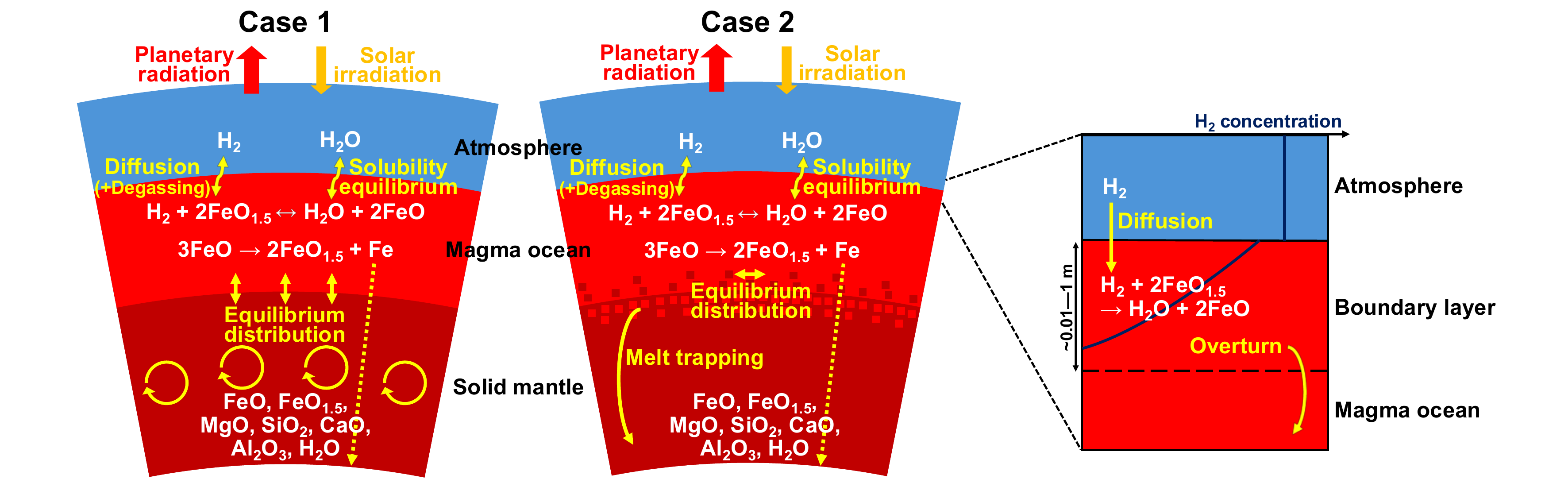}
	\caption{Schematic overview of the coupled atmosphere-interior evolution model. This model considers the diffusion of atmospheric H$_2$ into the magma ocean, redox reactions in the magma ocean, partitioning of oxide components and H$_2$O among each phase, and the cooling through the thermal planetary radiation, with Fe produced by Fe$^{2+}$ disproportionation removed into the core as it forms. Cases 1 and 2 represent two end-member treatments of material exchange between the magma ocean and solid mantle, introduced to account for uncertainty in the efficiency of material exchange between these reservoirs. Case 1 assumes efficient material exchange and equilibrium partitioning of oxide components and H$_2$O between the bulk magma ocean and solid mantle with uniform chemical compositions. Case 2 assumes inefficient material exchange, in which melt is efficiently trapped in the rheological transition region without subsequent material mixing in the solid mantle. The melt-side boundary layer at the interface between the magma ocean and atmosphere is considered to evaluate the H$_2$ diffusion rate into the magma ocean, with its overturn frequency determined by buoyancy-driven convective instability.}
	\label{fig:1}
\end{figure*}
	
\section{Model description} \label{sec:model}
	We investigate the coupled evolution of Earth's proto-atmosphere, composed of H$_2$ and H$_2$O, and the interior just after a giant impact event. A schematic overview of our model is shown in Figure~\ref{fig:1}. Initially, the atmosphere is assumed to be H$_2$-rich, reflecting the reduced condition during the core formation phase, while the magma ocean is oxidized by the disproportionation of Fe$^{2+}$ and efficient segregation of Fe into the core. 
	
	In this model, the compositional evolution of the magma ocean and atmosphere is primarily driven by the diffusion of atmospheric H$_2$ into the magma ocean and subsequent redox reactions within melt. The melt-side boundary layer at the interface between the magma ocean and atmosphere is considered to evaluate the H$_2$ diffusion rate into the magma ocean, with its overturn frequency determined by buoyancy-driven convective instability. The atmospheric-side boundary layer is not explicitly resolved, because gas transport is expected to be limited mainly by diffusion in the silicate melt, whose molecular diffusivity is orders of magnitude smaller than that in the atmosphere \citep{Olson2018,Olson2019}. 
	
	The partitioning of oxide components between the magma ocean and solid mantle is included to evaluate the magma ocean chemical composition under two different conditions (Case 1 and Case 2) since the efficiency of material exchange between the magma ocean and mantle is uncertain. Case 1 assumes efficient material exchange and equilibrium partitioning of oxide components and H$_2$O between the bulk magma ocean and solid mantle with uniform chemical compositions. Case 2 assumes inefficient material exchange, in which melt is efficiently trapped in the rheological transition region without subsequent material mixing in the solid mantle. The partitioning of H$_2$O among the mantle, magma ocean, and atmosphere is likewise treated, considering the two cases for the distribution between the mantle and magma ocean, together with solubility equilibrium between the magma ocean and atmosphere. Although atmospheric H$_2$ diffusion into the magma ocean is explicitly considered, H$_2$ is not treated as a stored volatile component partitioned into the magma ocean and solid mantle, given its low solubility in silicate melt compared to H$_2$O and its conversion to H$_2$O under the oxidizing conditions of the magma ocean. When redox reactions with atmospheric H$_2$ reduce the magma ocean sufficiently for Fe to become stable, we assume that any produced Fe is efficiently and continuously removed from the base of the magma ocean into the core as it forms, rather than remaining in the magma ocean as a redox buffer, while Fe$^{2+}$ disproportionation adjusts the Fe$^{2+}$ and Fe$^{3+}$ abundances to maintain the oxygen fugacity at the base of the magma ocean at the Fe-FeO coexistence condition. 
	
	The planetary thermal evolution proceeds simultaneously, with the magma ocean heat flux determined by the net planetary radiation flux. The net planetary radiation flux is derived from radiative transfer calculations of planetary and solar radiation in the atmosphere. The calculation of the coupled evolution is terminated when the base of the magma ocean reaches the surface. 
	
	Below, we describe the structure and composition of the magma ocean and solidified mantle in Section 2.1, atmospheric structure and radiative transfer in Section 2.2, and the diffusion of atmospheric H$_{2}$ through the boundary layer and redox reactions in the magma ocean in Section 2.3. Details of the thermal evolution calculation are described in the Appendix and \citet{Yoshida2025}.

\begin{figure*}
\gridline{\fig{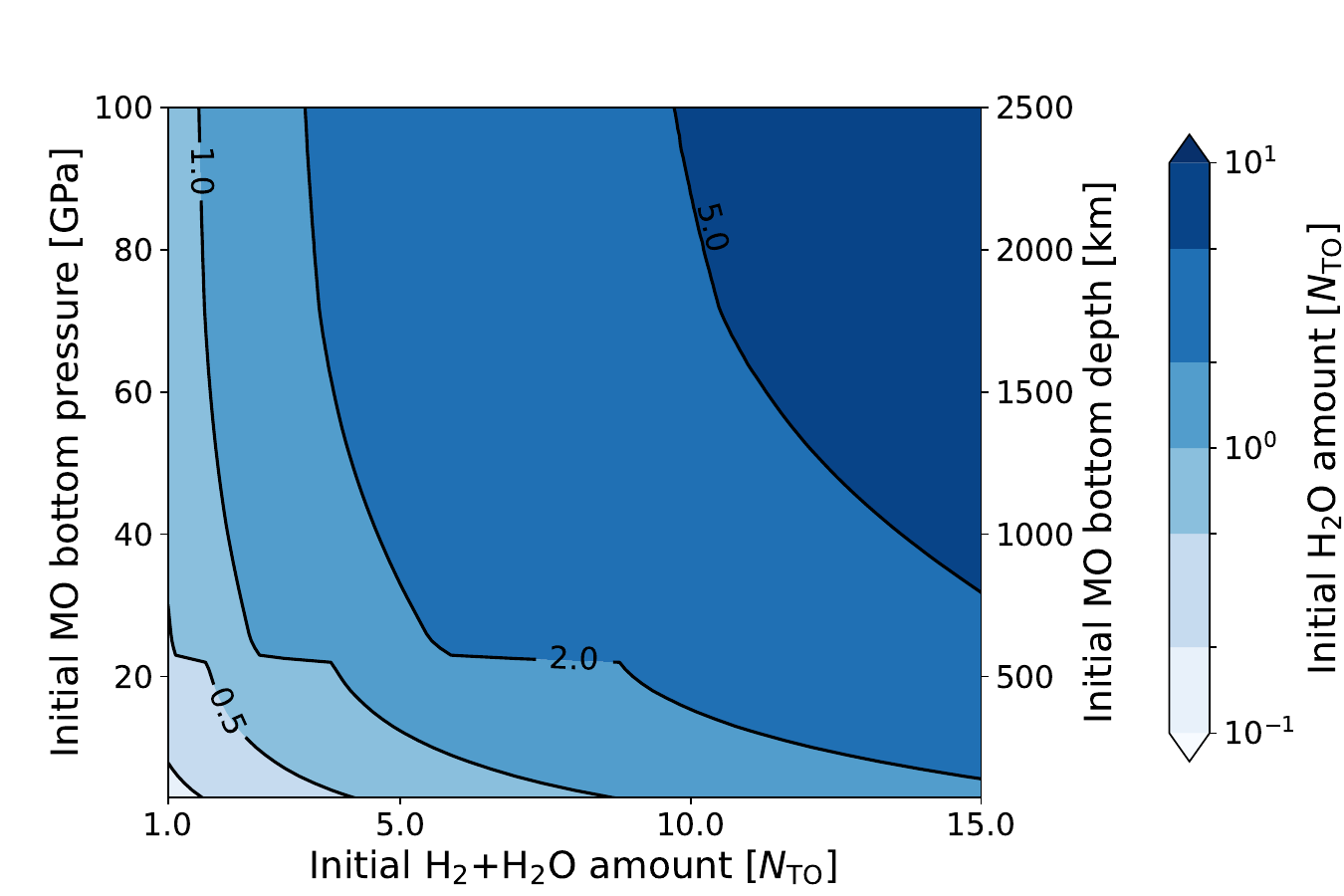}{0.45\textwidth}{(a)}
		\fig{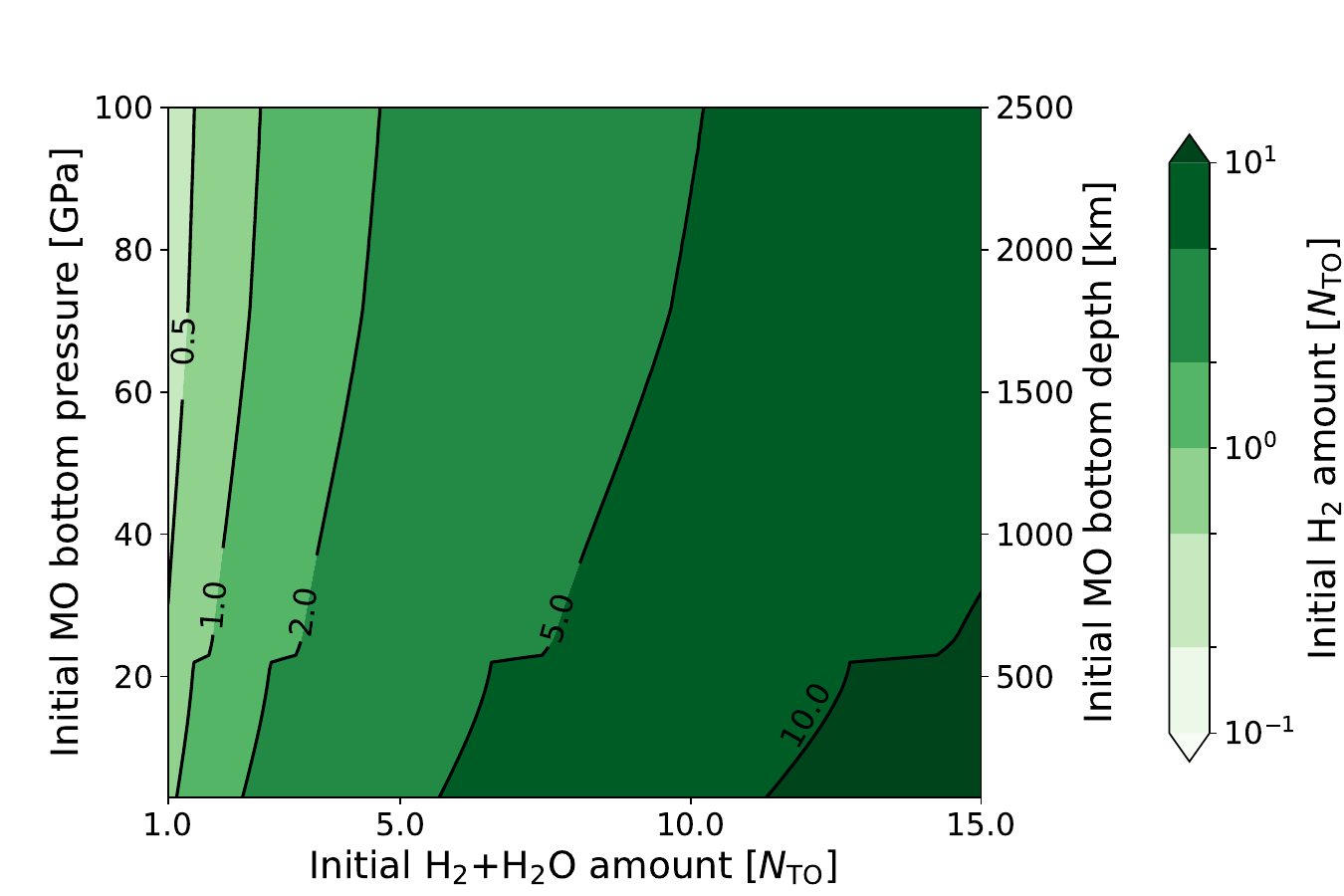}{0.45\textwidth}{(b)}
          	}
\caption{Initial amounts of H$_2$O (a) and H$_2$ (b) as a function of the initial H$_2$+H$_2$O amount and the initial bottom pressure of the magma ocean in Case 1 described in Section 2.1. The amounts are normalized by the hydrogen molecular number in the present-day terrestrial seawater ($N_{\rm TO}$).}
\label{fig:2}
\end{figure*}
	
\subsection{Structure and composition of the magma ocean and mantle}
	We consider a spherically symmetric, hydrostatic magma ocean overlying a solidified mantle, following \citet{Yoshida2025}. The temperature profile is taken to follow an adiabatic gradient supposing vigorous convection. The melt fraction $\phi$ at each depth is parameterized based on the local temperature relative to the solidus and liquidus temperature ($=(T-T_{\rm sol})/(T_{\rm liq}-T_{\rm sol})$), where $T$, $T_{\rm sol}$, and $T_{\rm liq}$ are the temperature, solidus temperature, and liquidus temperature, respectively \citep[e.g.,][]{Abe1997,Solomatov2007,Lebrun2013}. The base of the magma ocean is defined as the depth at which the melt fraction reaches the value at the rheological transition (=0.4) \citep{Abe1997,Lebrun2013,Schaefer2024}. The initial depth of the magma ocean, corresponding to the characteristic depth of metal-silicate equilibration, is treated as a free parameter with a basal pressure ranging from 1 to 100 GPa. Details of the interior structure are described in Appendix C.
	
	We consider FeO, FeO$_{1.5}$, MgO, SiO$_2$, CaO, and Al$_2$O$_3$ as the primary oxide components of the magma ocean and the mantle, following \citet{Schaefer2024}. The total abundances of the oxides except FeO and FeO$_{1.5}$ in the magma ocean and mantle are taken from the bulk silicate Earth (BSE) composition reported by \citet{Palme2014}. The initial total amount of FeO+FeO$_{1.5}$ is set equal to that of the BSE (=8.1wt\%). The initial ratio of FeO$_{1.5}$ to FeO is described in the final paragraph of this subsection. The partitioning of H$_2$O into the magma ocean and mantle is also considered. Although H$_2$ diffusion into the magma ocean is explicitly considered as described in Section 2.3, H$_2$ is not treated as a stored volatile component partitioned into the magma ocean and solid mantle, given its low solubility in silicate melt compared to H$_2$O and its conversion to H$_2$O under the oxidizing conditions of the magma ocean. The initial amount and distribution of the volatiles are described in Section 2.2. The magma ocean is supposed to be chemically homogeneous due to vigorous convection \citep{Lebrun2013}, except within the boundary layer at the interface between the magma ocean and atmosphere as described in Section 2.3.  
	
	We consider the partitioning of the oxide components and H$_2$O among coexisting phases using two endmember approaches to account for uncertainty in the efficiency of material exchange between the magma ocean and solid mantle. In Case 1, supposing the efficient interaction between the bulk magma ocean and the bulk solidified mantle, the concentrations of component $i$ except H$_2$O in the melt and in the crystals are described using the mass-weighted average partition coefficient and taking into account the mass conservation, 
	\begin{equation}
		\frac{X_{\mathrm{m},i}^{\rm crys}}{X_{\mathrm{m},i}^{\rm melt}}=K_{\mathrm{D},{i}}=\sum_{j}W^{\mathrm{crys,T}}_{j}K_{\mathrm{D},{i}}^{j},
	\end{equation}
	\begin{equation}
		X_{\mathrm{m},i}^{\rm melt}M_{\rm melt}^{\mathrm{T}} +X_{\mathrm{m},i}^{\rm crys}M_{\rm crys}^{\mathrm{T}}=M_{i}^{\mathrm{T}},
	\end{equation}
	where $X_{\mathrm{m},i}^{\rm crys}$ and $X_{\mathrm{m},i}^{\rm melt}$ are the mass fractions of component $i$ in the crystalline phases and in the melt, respectively, $K_{\mathrm{D},{i}}^{j}$ is the partition coefficient of component $i$ between melt and mineral phase $j$, $W_{j}^{\rm crys,T}$ is the mass fraction of mineral phase $j$ in the total crystalized cumulates, $K_{\mathrm{D},{i}}$ is the average partition coefficient of species $i$, $M_{\rm melt}^{\mathrm{T}}$ and $M_{\rm crys}^{\mathrm{T}}$ are the total mass of melt and crystallized cumulates, and $M_{i}^{\rm T}$ is the total mass of component $i$ in the magma ocean and solid mantle. We consider olivine, spinel, clinopyroxene, orthopyroxene, garnet, majorite, wadsleyite, ringwoodite, bridgmanite, and magnesiow\"{u}stite as crystallizing mineral phases \citep{Schaefer2024}. The crystallization sequence is taken from \citet{Elkins2008}, and mineral-melt partition coefficients from \citet{Schaefer2024}. From these equations, $X_{\mathrm{m},i}^{\rm melt}$ and $X_{\mathrm{m},i}^{\rm crys}$ are described as follows:
	\begin{equation}
		X_{\mathrm{m},i}^{\rm melt}=\frac{M_{i}^{\rm T}}{M_{\rm melt}^{\mathrm{T}}+K_{\mathrm{D},i}M_{\rm crys}^{\mathrm{T}}},
	\end{equation}
	\begin{equation}
		X_{\mathrm{m},i}^{\rm crys}=\frac{K_{\mathrm{D},i}M_{i}^{\rm T}}{M_{\rm melt}^{\mathrm{T}}+K_{\mathrm{D},i}M_{\rm crys}^{\mathrm{T}}}.
	\end{equation}
	For H$_2$O, the solubility equilibrium between the magma ocean and the atmosphere, as well as the equilibrium distribution between the magma ocean and the solidified mantle, is also taken into account using the solubility law given by
	\begin{equation}
		P_{\rm H_{2}O}=\left(\frac{X^{\rm melt}_{\mathrm{m},\mathrm{H_{2}O}}}{\alpha_{\rm H_{2}O}}\right)^{\beta_{\rm H_{2}O}},
	\end{equation}
	where $P_{\rm H_{2}O}$ is the atmospheric partial pressure of H$_2$O in Pa and $\alpha_{\rm H_{2}O}(=6.8\times 10^{-8}\,\mathrm{Pa}^{-1/\beta_{\rm H_{2}O}})$ and $\beta_{\rm H_{2}O}(=1.428)$ are the Henrian fit coefficients of H$_2$O \citep{Lebrun2013}. The mass balance of H$_2$O is given by
	\begin{equation}
		X^{\rm melt}_{\mathrm{m},\mathrm{H_{2}O}}M_{\rm melt}^{\rm T}+K_{\rm D,H_{2}O} X^{\rm melt}_{\mathrm{m},\mathrm{H_{2}O}}M_{\rm crys}^{\rm T}+\frac{4\pi R_{\rm p}^{2}}{g}\frac{m_{\rm H_{2}O}}{\bar{m}}P_{\rm H_{2}O}=M_{\rm H_{2}O}^{\rm T},
	\end{equation}
	where $R_{\rm p}$ is the planetary radius, $g$ is the gravitational acceleration, $m_{\rm H_{2}O}$ is the molecular mass of H$_2$O, and $\bar{m}$ is the atmospheric mean molecular mass. The mineral-melt partition coefficient of H$_2$O is taken from \citet{Lebrun2013}.
	
	In Case 2, we assume that melts in the rheological transition region are efficiently trapped in the solidified mantle without the following material exchange between the magma ocean and solidified mantle \citep{Miyazaki2022a,Miyazaki2022b}. In this case, equilibrium partitioning between melt and crystals is assumed to be established locally within the magma ocean. $K_{\mathrm{D},i}$, $X_{\mathrm{m},i}^{\rm melt}$, and $X_{\mathrm{m},i}^{\rm crys}$ are described as follows:
	\begin{equation}
		K_{\mathrm{D},{i}}=\sum_{j}W^{\mathrm{crys,MO}}_{j}K_{\mathrm{D},{i}}^{j},
	\end{equation}
	\begin{equation}
		X_{\mathrm{m},i}^{\rm melt}M_{\rm melt}^{\mathrm{MO}} +X_{\mathrm{m},i}^{\rm crys}M_{\rm crys}^{\mathrm{MO}}=M_{i}^{\mathrm{MO}},
	\end{equation}
	\begin{equation}
		X_{\mathrm{m},i}^{\rm melt}=\frac{M_{i}^{\rm MO}}{M_{\rm melt}^{\mathrm{MO}}+K_{\mathrm{D},i}M_{\rm crys}^{\mathrm{MO}}},
	\end{equation}
	\begin{equation}
		X_{\mathrm{m},i}^{\rm crys}=\frac{K_{\mathrm{D},i}M_{i}^{\rm MO}}{M_{\rm melt}^{\mathrm{MO}}+K_{\mathrm{D},i}M_{\rm crys}^{\mathrm{MO}}},
	\end{equation}
	where $W^{\mathrm{crys,MO}}_{j}$ is the mass fraction of mineral phase $j$ in the crystals in the magma ocean, $M_{\rm melt}^{\mathrm{MO}}$ and $M_{\rm crys}^{\mathrm{MO}}$ are the mass of melt and crystals in the magma ocean, and $M_{i}^{\rm MO}$ is the mass of component $i$ in the magma ocean. For H$_2$O, the solubility equilibrium between the magma ocean and atmosphere, as well as the equilibrium partitioning between melt and crystals in the magma ocean, is considered as in Case 1, and the mass balance is given by
	\begin{equation}
		X^{\rm melt}_{\mathrm{m},\mathrm{H_{2}O}}M_{\rm melt}^{\rm MO}+K_{\rm D,H_{2}O} X^{\rm melt}_{\mathrm{m},\mathrm{H_{2}O}} M_{\rm crys}^{\rm MO}+\frac{4\pi R_{\rm p}^{2}}{g}\frac{m_{\rm H_{2}O}}{\bar{m}}P_{\rm H_{2}O}=M_{\rm H_{2}O}^{\rm MO}.
	\end{equation}
	The evolution of the total amount of each component retained in the magma ocean (and the atmosphere for H$_2$O) during solidification is given by
	\begin{equation}
		dM_{i}^{\mathrm{MO+atm}}=[\phi_{\rm crit} X_{\mathrm{m},i}^{\rm melt}+(1-\phi_{\rm crit})X_{\mathrm{m},i}^{\rm crys}]dM_{\rm MO},
	\end{equation}
	where $M_{i}^{\mathrm{MO+atm}}$ is the total mass of component $i$ in the magma ocean and atmosphere, $M_{\rm MO}$ is the mass of the magma ocean, and $\phi_{\rm crit}(=0.4)$ is the melt fraction at the rheological transition.  
	
	The oxygen fugacity of the magma ocean is calculated as a function of temperature, pressure, and magma ocean composition using the recalibrated equation of \citet{Hirschmann2022}:
	\begin{align}
		a\mathrm{log}f_{\mathrm{O_2}}
		    &= \mathrm{log}\left(\frac{X^{\rm melt}_{\mathrm{FeO_{1.5}}}}
		    {X^{\rm melt}_{\mathrm{FeO}}}\right)
		    - b - \frac{c}{T}
		    + \frac{\Delta C_{p}}{R\mathrm{ln}(10)}
		    \left[1-\frac{T_0}{T}-\mathrm{ln}\left(\frac{T}{T_0}\right)\right]
		    \notag \\
		    &\quad + \frac{1}{RT\mathrm{ln}(10)}
		    \int_{P_0}^{P}\Delta V\,dP
		    - \frac{1}{T}\left(
		    Y_1 X^{\rm melt}_{\mathrm{SiO_2}}
		    + Y_2 X^{\rm melt}_{\mathrm{TiO_2}}
		    \right. \notag \\
		    &\qquad\left.
		    + Y_3 X^{\rm melt}_{\mathrm{MgO}}
		    + Y_4 X^{\rm melt}_{\mathrm{CaO}}
		    + Y_5 X^{\rm melt}_{\mathrm{NaO_{0.5}}}
		    + Y_6 X^{\rm melt}_{\mathrm{KO_{0.5}}}
		    \right. \notag \\
		    &\qquad\left.
		    + Y_7 X^{\rm melt}_{\mathrm{PO_{2.5}}}
		    + Y_8 X^{\rm melt}_{\mathrm{SiO_2}}X^{\rm melt}_{\mathrm{AlO_{1.5}}}
		    + Y_9 X^{\rm melt}_{\mathrm{SiO_2}}X^{\rm melt}_{\mathrm{MgO}}
		    \right),
	\end{align}
	where $f_{\rm O_{2}}$ is the oxygen fugacity, $X_{i}^{\rm melt}$ is the molar fraction of species $i$ in the melt, $R$ is the gas constant, $T$ is the temperature, $P_{0}=10^{5}\,\mathrm{Pa}$ is the reference pressure, $\Delta V$ is the difference in molar volume between FeO and FeO$_{1.5}$, and $a$, $b$, $c$, $\Delta C_{p}$, $T_0$, and $Y_{i}$ are the fitting parameters given in Table 2 of \citet{Hirschmann2022}. The molar volumes are estimated using the EOS calibrated by \citet{Zhang2024}, which is based on the EOS developed by \citet{Deng2020}. We set $X_{\rm TiO_2}^{\rm melt}=X_{\rm NaO_{0.5}}^{\rm melt}=X_{\rm KO_{0.5}}^{\rm melt}=X_{\rm PO_{2.5}}^{\rm melt}=0$.
	
	The initial FeO$_{1.5}$/FeO ratio is set such that the oxygen fugacity at the base of the magma ocean corresponds to the value at which iron oxides and metallic iron coexist: $f_{\rm O_{2}}=(a_{\rm FeO}/a_{\rm Fe})^{2}f_{\rm O_{2}}^{\rm IW}=(\gamma_{\rm FeO}X_{\rm FeO}/a_{\rm Fe})^{2}f_{\rm O_{2}}^{\rm IW}$, where $f_{\rm O_{2}}^{\rm IW}$ is the oxygen fugacity at IW buffer \citep{Hirschmann2021}, and $a_{\rm FeO}$, $\gamma_{\rm FeO}(=1.55)$, and $a_{\rm Fe}(=0.75)$ are the activity of FeO, activity coefficient of FeO, and activity of Fe, respectively \citep{Hirschmann2022}. Redox reactions with atmospheric H$_2$ can lower the oxygen fugacity of the magma ocean to conditions where Fe becomes stable. When Fe is produced under such reducing conditions, we assume that it is efficiently and continuously removed from the base of the magma ocean into the core as it forms, and therefore does not remain in the magma ocean as a redox buffer. Accordingly, the Fe$^{2+}$ and Fe$^{3+}$ abundances are adjusted by Fe$^{2+}$ disproportionation so that the oxygen fugacity at the base of the magma ocean is maintained at the Fe-FeO coexistence condition, $f_{\mathrm{O_{2}}}=(a_{\mathrm{FeO}}/a_{\mathrm{Fe}})^2 f_{\mathrm{IW}}$.

\subsection{Structure, composition, and radiative properties of the proto-atmosphere}
	We suppose that the proto-atmosphere is composed of H$_2$ and H$_2$O. Other components, such as C- and N-bearing species, are neglected for simplicity. The initial atmospheric H$_2$/H$_2$O ratio is determined by thermochemical equilibrium under the oxygen fugacity $f_{\rm O_{2}}=(a_{\rm FeO}/a_{\rm Fe})^{2}f_{\rm O_{2}}^{\rm IW}$, which reflects the reduced environment during the impact degassing and core formation phase \citep{Kuramoto1996,Frost2008,Hirschmann2012}. This typical molar H$_2$/H$_2$O ratio is $\sim 10$, with a little dependence on temperature and pressure. Here, the chemical equilibrium constant is calculated from the standard chemical potentials of the species involved, as described in detail in Section 2.3. The initial condition also satisfies the H$_2$O partitioning between the atmosphere and the interior described in Section 2.1. In Case 1, the solubility equilibrium between the magma ocean and the atmosphere, as well as the equilibrium distribution between the magma ocean and the solidified mantle, is assumed. In Case 2, the solubility equilibrium between the magma ocean and the atmosphere is also assumed, but partitioning into the solid mantle is initially neglected. The total amount of H$_2$+H$_2$O is treated as a free parameter. Figure~\ref{fig:2} shows the initial amount of H$_2$O and H$_2$ as a function of the initial total amount and the initial bottom pressure of the magma ocean in Case 1. Note that the initial amount in Case 2 is nearly identical to that in Case 1. 
	
	We perform radiative transfer calculations in a plane-parallel atmosphere on the two-stream approximation for the planetary and solar radiation separately to derive the outgoing planetary radiation flux and the planetary albedo as functions of atmospheric mass, composition, and surface temperature \citep{Yoshida2025}. As radiative absorption sources, we include H$_2$O line absorption, H$_2$O continuum absorption using the MT\_CKD model \citep{Mlawer2023}, including both self and foreign continuum components, and H$_2$-H$_2$ collision-induced absorption, while opacity from other volatile species is not included. Rayleigh scattering by H$_2$O and H$_2$ is also considered.  The planetary cooling rate is supposed to be along with the net energy balance of the incoming and outgoing radiation at the top of the atmosphere. The temperature profile in the lower part of the atmosphere follows an adiabatic lapse rate, while that in the upper part is given by the skin temperature, which is the asymptotic temperature at high altitudes of an upper atmosphere that is optically thin in the thermal-IR and transparent to shortwave radiation, following \citet{Yoshida2025}. Details of the atmospheric structure are provided in Appendix A and \citet{Yoshida2025}. Further details of the radiative transfer calculations are provided in Appendix B and \citet{Yoshida2025}. 

\subsection{Diffusive transport and redox reactions in the magma ocean boundary layer}
	To evaluate the diffusion flux of atmospheric H$_2$ into the magma ocean through the melt-side boundary layer at the interface between the magma ocean and atmosphere, we solve the diffusion-reaction equation:
	\begin{equation}
		\frac{\partial n_{i}(z,t)}{\partial t}=\frac{\partial}{\partial z}\left(D_{i}(z,t)\frac{\partial n_{i}(z,t)}{\partial z}\right)+S_{i}(z,t),
	\end{equation}
	where $t$ is time, $z$ is depth from the planetary surface, and $n_{i}(z,t)$, $D_{i}(z,t)$, and $S_{i}(z,t)$ are the number density, diffusion coefficient, and chemical production/loss rate, respectively, of species $i$ ($i=$H$_{2}$, H$_2$O, Fe$^{3+}$, Fe$^{2+}$) in the melt at depth $z$ and time $t$. The upper boundary is set at the planetary surface, where the H$_2$ density is specified from the solubility equilibrium between the melt and the atmosphere. The solubility law for H$_2$ and the density of H$_2$ at the upper boundary are given by
	\begin{equation}
		P_{\rm H_{2}}=\left(\frac{X^{\rm melt}_{\mathrm{m},\mathrm{H_{2}}}}{\alpha_{\rm H_{2}}}\right)^{\beta_{\rm H_{2}}},
	\end{equation}
	\begin{equation}
		\rho_{\rm H_2}(0,t) = \rho_{\rm melt} \alpha_{\rm H_2} P_{\rm H_2}^{1/\beta_{\rm H_2}},
	\end{equation}
	where $P_{\rm H_{2}}$ is the atmospheric partial pressure of H$_2$ in Pa, $\alpha_{\rm H_{2}}(=2.572 \times 10^{-12}\,\mathrm{Pa^{-1}})$ and $\beta_{\rm H_{2}}(=1.0)$ are the Henrian fit coefficients of H$_2$ \citep{Lichtenberg2021}, $\rho_{\rm H_2}$ ($=n_{\rm H_2}m_{\rm H_2}$) is the mass density of H$_2$ in the melt, $m_{\rm H_2}$ is the molecular mass of H$_2$, and $\rho_{\rm melt}$ ($=4.0\times 10^{3}$ kg/m$^{3}$) is the melt density \citep{Lebrun2013}. The lower boundary is set at a depth sufficiently greater than the diffusion distances of H$_2$ and H$_2$O. The initial total abundance of FeO+FeO$_{1.5}$ is set equal to that of the BSE (=8.1wt\%), and the initial and lower-boundary Fe$^{3+}$/Fe$^{2+}$ ratio is treated as a free parameter. The initial H$_2$O abundance in the boundary layer is set to zero, and zero-flux boundary conditions are imposed for H$_2$O at both the upper and lower boundaries.
	
	The diffusion coefficients in the melt are given by
	\begin{equation}
		D_{\rm H_{2}}=\mathrm{exp}\left(-16.471-\frac{5.363\times 10^{3}}{T_{\rm s}} \right),
	\end{equation}
	\begin{equation}
		\begin{aligned}
			D_{\rm H_{2}O}
			    &= \mathrm{exp}\left(
			    -14.26
			    -37.26X_{\rm H_{2}O}^{\rm melt}(z,t)
			    \right. \\
			    &\qquad\left.
			    -\frac{1.294\times 10^{4}
			    -7.588\times 10^{4}X_{\rm H_{2}O}^{\rm melt}(z,t)}
			    {T_{\rm s}}
			    \right),
		\end{aligned}
	\end{equation}
	\begin{equation}
		D_{\rm Fe^{3+}}=D_{\rm Fe^{2+}}=\mathrm{exp}\left(-7.83-\frac{2.442\times 10^{4}}{T_{\rm s}} \right),
	\end{equation}
	where the diffusion coefficients are expressed in m$^2$/s \citep{Zhang2010a,Zhang2010b} and $T_{\rm s}$ is the surface temperature. Because available experimental data are insufficient to independently constrain the diffusivities of Fe$^{2+}$ and Fe$^{3+}$, we assume the same effective diffusion coefficient for both oxidation states, adopting the representative Fe tracer diffusivity of \citet{Zhang2010b}. In addition, these fitting formulas are valid up to approximately 1700 K due to the limit of experimental data; at higher temperatures, the diffusion coefficients are set to their values at 1700 K. 
	
	The considered chemical process is the redox reaction between H$_2$ and Fe$^{3+}$:
	\begin{equation}
		\mathrm{H_2}+2\mathrm{Fe^{3+}} \to 2\mathrm{Fe^{2+}}+2\mathrm{H^{+}} \quad (\mathrm{H_{2}}+\mathrm{2FeO_{1.5}} \to \mathrm{2FeO}+\mathrm{H_{2}O}).
	\end{equation}
	 The chemical production/loss rates in units of m$^{-3}$ s$^{-1}$ are described as follows \citep{Barde2016}:
	 \begin{equation}
		\begin{aligned}
			-S_{\rm Fe^{3+}}
			    &= +S_{\rm Fe^{2+}}
			    = -2S_{\rm H_{2}}
			    = +2S_{\rm H_{2}O} \\
			    &= 4.2\times 10^{-27}
			    \mathrm{exp}\left(-\frac{5.68\times 10^{3}}{T_{\rm s}}\right)
			    n_{\rm H_{2}}^{\rm equiv}(z,t)
			    n_{\rm Fe^{3+}}(z,t),
		\end{aligned}
	\end{equation}
	where $n_{\rm H_2}^{\rm equiv}(z,t)$ is the equivalent gas-phase number density of H$_2$. Because \citet{Barde2016} derived a relationship between the reaction rate and the surrounding gas concentration, the density of H$_2$ in the melt needs to be converted to the corresponding gas density. We assume that the equivalent number density is given by the gas density that is in solubility equilibrium with the melt containing H$_2$ as follows:
	\begin{equation}
		n_{\rm H_2}^{\rm equiv}(z,t)=\frac{1}{k_{\rm B}T_{\rm s}}\left(\frac{1}{\alpha_{\rm H_{2}}} \frac{\rho_{\rm H_2}(z,t)}{\rho_{\rm melt}}\right)^{\beta_{\rm H_{2}}},
	\end{equation}
	where $k_{\rm B}$ is the Boltzmann constant.
	
	The forward redox reaction in Equation (20) proceeds while the local H$_2$O/H$_2$ ratio is below its equilibrium value, and the chemical source terms $S_i$ are set to zero once local redox equilibrium is reached. The equilibrium condition in the melt is derived below from the reaction Gibbs energy. The chemical potentials in the melt, expressed relative to the gas-phase standard state, and the reaction Gibbs energy are given by
	\begin{equation}
		\mu_{\rm H_2}^{\rm melt}(z,t) = \mu_{\rm H_2}^{\circ}(T_{\rm s}) + RT_{\rm s}\mathrm{ln}\left(\frac{f_{\rm H_2}^{\rm equiv}(z,t)}{f^{\circ}}\right),
	\end{equation}
	\begin{equation}
		\mu_{\rm H_{2}O}^{\rm melt}(z,t) = \mu_{\rm H_{2}O}^{\circ}(T_{\rm s}) + RT_{\rm s}\mathrm{ln}\left(\frac{f_{\rm H_{2}O}^{\rm equiv}(z,t)}{f^{\circ}}\right),
	\end{equation}
	\begin{equation}
		\mu_{\rm O_{2}}^{\rm melt}(z,t) = \mu_{\rm O_{2}}^{\circ}(T_{\rm s}) + RT_{\rm s}\mathrm{ln}\left(\frac{f_{\rm O_{2}}(z,t)}{f^{\circ}}\right),
	\end{equation}
	\begin{equation}
		\Delta_{\rm r}G_{\rm melt}(z,t) = \mu_{\rm H_{2}O}^{\rm melt}(z,t) - \mu_{\rm H_2}^{\rm melt}(z,t) - \frac{1}{2}\mu_{\rm O_{2}}^{\rm melt}(z,t),
	\end{equation}
	where $\mu_{i}^{\rm melt}(z,t)$ is the chemical potential of dissolved species $i$ ($i=$H$_2$, H$_2$O) in the melt, $\mu_{\rm O_2}^{\rm melt}(z,t)$ is the oxygen chemical potential corresponding to the redox state of the melt, expressed in terms of the oxygen fugacity $f_{\rm O_2}(z,t)$, $\mu_{i}^{\circ}(T_{\rm s})$ is the standard chemical potential of gaseous species $i$ at the standard fugacity $f^{\circ}=1\,\mathrm{bar}$ and at the surface temperature, $\Delta_{\rm r} G_{\rm melt}(z,t)$ is the reaction Gibbs energy in the melt, and $f_{i}^{\rm equiv}(z,t)$ is the equivalent gas-phase fugacity, defined as the fugacity of a hypothetical gas in solubility equilibrium with the melt. At gas--melt equilibrium, the chemical potential of a dissolved species is equal to that of the corresponding gas phase. Therefore, the equivalent fugacity corresponding to a given dissolved concentration is obtained from the solubility law (Equations (5) and (15) for H$_2$O and H$_2$, respectively) as
	\begin{equation}
		f_{i}^{\rm equiv}(z,t)=\left(\frac{1}{\alpha_{i}} \frac{\rho_{i}(z,t)}{\rho_{\rm melt}}\right)^{\beta_{i}},
	\end{equation}
	where $\rho_{i}(z,t) (=n_{i}m_{i})$ is the density of species $i$ in the melt. From the chemical equilibrium condition, $\Delta_{\rm r} G_{\rm melt}=0$, the equilibrium equivalent-fugacity ratio is given by
	\begin{equation}
		\frac{f_{\rm H_{2}O}^{\rm equiv}}{f_{\rm H_2}^{\rm equiv}} = K_{\rm eq}(T_{\rm s}) \left( \frac{f_{\rm O_2}}{f^{\circ}} \right)^{1/2},
	\end{equation}
	\begin{equation}
		K_{\rm eq}(T_{\rm s})=\mathrm{exp}\left(-\frac{\mu_{\rm H_{2}O}^{\circ}(T_{\rm s})-\mu_{\rm H_{2}}^{\circ}(T_{\rm s})-\mu_{\rm O_{2}}^{\circ}(T_{\rm s})/2}{RT_{\rm s}}\right),
	\end{equation}
	where $K_{\rm eq}(T_{\rm s})$ is the chemical equilibrium constant at $T_{\rm s}$. Using Equation (27), the equilibrium fugacity ratio is converted to the corresponding H$_2$O/H$_2$ density ratio in the melt. Accordingly, the forward redox reaction in Equation (20) proceeds when
	\begin{equation}
		\begin{aligned}
			\frac{\rho_{\rm H_2O}(z,t)}{\rho_{\rm H_2}(z,t)}
			    &< \frac{\alpha_{\rm H_2O}}{\alpha_{\rm H_2}}
			    \left[
			    K_{\rm eq}(T_{\rm s})
			    \left(\frac{f_{\rm O_2}(z,t)}{f^{\circ}}\right)^{1/2}
			    \right]^{1/\beta_{\rm H_2O}} \\
			    &\qquad\times
			    \left(
			    \frac{1}{\alpha_{\rm H_2}}
			    \frac{\rho_{\rm H_2}(z,t)}{\rho_{\rm melt}}
			    \right)^{\beta_{\rm H_2}/\beta_{\rm H_2O}-1}.
		\end{aligned}
	\end{equation}
	The right-hand side of this inequality represents the equilibrium H$_2$O/H$_2$ density ratio in the melt. This equilibrium density ratio differs from the corresponding fugacity ratio because the conversion from fugacity to dissolved concentration depends on the species-specific solubility in the melt. The forward reaction ceases and the chemical source term $S_{i}$ are set to zero when equality is reached. The data of the standard chemical potential are taken from the NIST Chemistry WebBook \citep{Chase1998}.

\begin{figure*}
\gridline{\fig{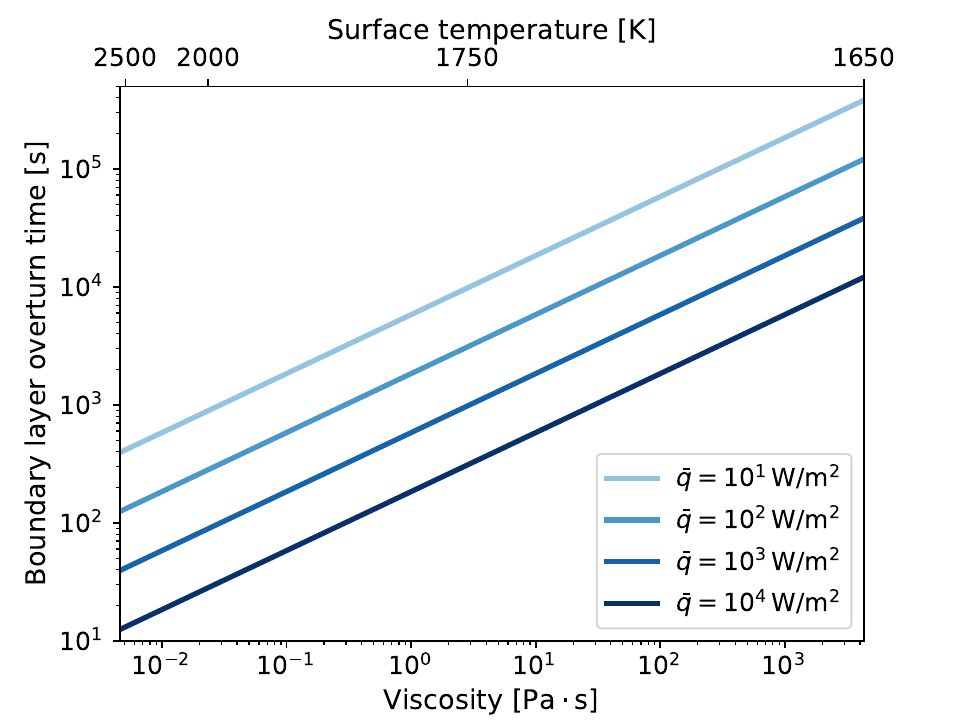}{0.45\textwidth}{(a)}
		\fig{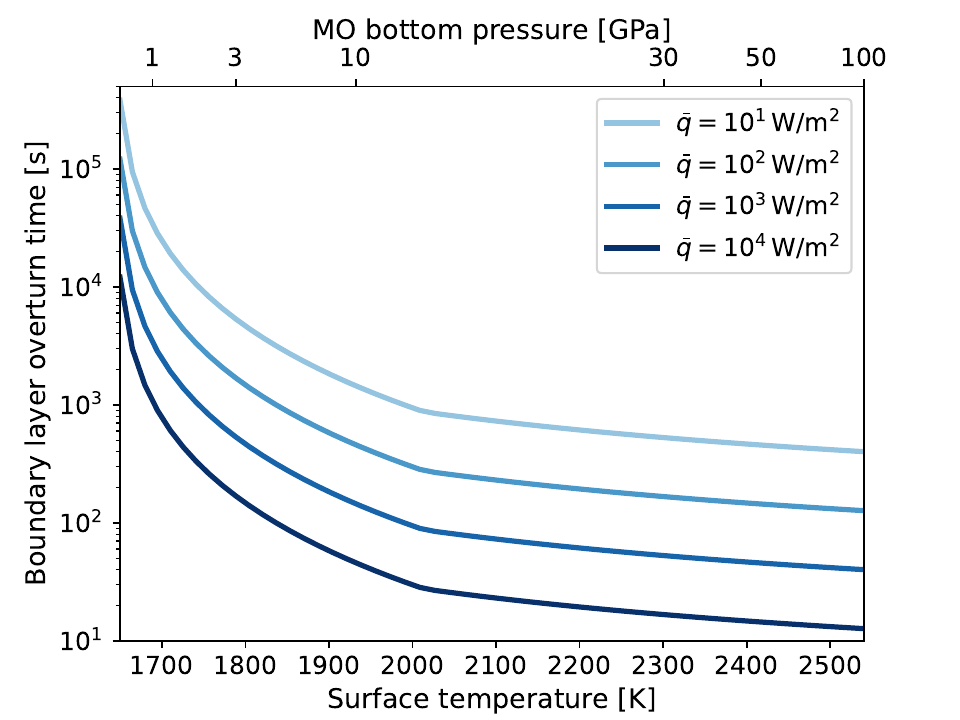}{0.45\textwidth}{(b)}
          	}
\caption{Overturn time of the boundary layer at the magma ocean--atmosphere interface, calculated from Equation (36), as a function of (a) magma ocean viscosity and the corresponding surface temperature and (b) surface temperature and the corresponding pressure at the bottom of the magma ocean. Colors indicate different internal heat fluxes.}
\label{fig:3}
\end{figure*}
	
	The mean H$_2$ diffusion flux $\bar{F}_{\rm H_2}$ over the boundary layer overturn time $t_{\rm c}$ is given by
	\begin{equation}
		\bar{F}_{\rm H_2}=\frac{1}{t_{\rm c}}\int_{0}^{t_{\rm c}}\left(-D_{\rm H_{2}}\left.\frac{\partial n_{\rm H_{2}}}{\partial z}\right|_{z=0}\right)dt.
	\end{equation}
	We estimate the overturn time using the transient boundary layer theory \citep{Turcotte2002}. In this framework, the cold, dense boundary layer formed by surface cooling is assumed to overturn when its Rayleigh number reaches a critical value, corresponding to the onset of buoyancy-driven convective instability. The critical Rayleigh number is given by
	\begin{equation}
		Ra_{\rm cr}=\frac{\rho_{\rm MO,s} \alpha_{T,\mathrm{s}} g (T_{1}-T_{0})z_{\rm T,cr}^{3}}{\kappa \eta_{\rm s}},
	\end{equation}
	where $Ra_{\rm cr}=1.1\times 10^{3}$ \citep{Schaefer2016}, $g$ is the gravitational acceleration, $\rho_{\rm MO,s}$, $\alpha_{T,\mathrm{s}}$, $\eta_{\rm s}$, and $\kappa$ are the density, thermal expansivity, viscosity, and thermal diffusivity of the magma ocean at the surface, respectively,  $z_{\rm T,cr}$ is the critical thickness of the boundary layer, $T_0$ and $T_1$ are the temperature at the top and bottom of the boundary layer, respectively. The value of $\kappa\,(=10^{-6}\,\mathrm{m^{2}/s})$ is taken from \citet{Lebrun2013}. The density and thermal expansivity of the magma ocean are described in Appendix C. $\eta_{\rm s}$ ($\mathrm{Pa\,s}$) is expressed by
	\begin{equation}
		\eta_{\rm s} = 0.00024\left(1-\frac{1-\phi_{\rm s}}{1-\phi_{\rm crit}}\right)^{-2.5}\mathrm{exp}\left(\frac{4600}{T_{\rm s}-1000}\right),
	\end{equation}	
	where $\phi_{\rm s}$ is the melt fraction at the surface \citep{Lebrun2013}. The critical thickness of the boundary layer ($z_{\rm T,cr}$) and the average internal heat flux ($\bar{q}$) during $t_{\rm c}$ are described as follows \citep{Turcotte2002}:
	\begin{equation}
		z_{\rm T,cr}=2.32(\kappa t_{\rm c})^{1/2},
	\end{equation}
	\begin{equation}
		\bar{q}=\frac{2k_{\rm s}(T_{1}-T_{0})}{\sqrt{\pi \kappa t_{\rm c}}},
	\end{equation}
	where $k_{\rm s}\,(=\kappa \rho_{\rm MO,s} C_{p,\mathrm{MO}})$ is the thermal conductivity at the surface and $C_{p,\mathrm{MO}}\,(=1000\,\mathrm{J/kg/K})$ is the specific heat of the magma ocean \citep{Lebrun2013}. We suppose that the internal heat flux is equal to the net outgoing planetary radiation flux at the top of the atmosphere. By imposing the condition that the Rayleigh number reaches its critical value ($Ra=Ra_{\rm cr}$), the boundary layer overturn time $t_{\rm c}$ is derived as:
	\begin{equation}
		t_{\rm c}=0.40\left(\frac{k_{\rm s} \eta_{\rm s}}{\pi^{1/2} \rho_{\rm MO,s} \alpha_{T,\mathrm{s}} g \kappa \bar{q}}Ra_{\rm cr}\right)^{1/2}.
	\end{equation}
	The dependence of the overturn time on surface temperature and internal heat flux derived from Equation (36) is shown in Figure~\ref{fig:3}. 
	
	From the equations above, we obtain the mean H$_2$ diffusion flux $\bar{F}_{\rm H_2}$ as a function of H$_2$ partial surface pressure, surface temperature, Fe$^{3+}$ concentration, and boundary layer overturn time. Any H$_2$ remaining in the boundary layer at the time of overturn is assumed to be transported into and mixed with the oxidized bulk magma ocean, where it is eventually converted to H$_2$O. Accordingly, the H$_2$O production and the associated production of FeO and consumption of FeO$_{1.5}$ are calculated from the mean H$_2$ diffusion flux based on the stoichiometry of Equation (20).
	
	H$_2$ diffusion into the magma ocean is allowed to proceed until redox equilibrium at the interface between the magma ocean and atmosphere is achieved. The reaction Gibbs energy and chemical potential of species $i$ governing the surface redox equilibrium are given by
	\begin{equation}
		\Delta_{\rm r} G_{\rm s}=\mu_{\mathrm{H_2O}}^{\rm atm,s}-\mu_{\mathrm{H_2}}^{\rm atm,s}-\frac{1}{2}\mu_{\mathrm{O_2}}^{\rm s},
	\end{equation}
	\begin{equation}
		\mu_{i}^{\rm atm,s}=\mu_i^\circ(T_{\rm s})+RT_{\rm s}\ln\left(\frac{f_{i}^{\rm atm,s}}{f^\circ}\right),
	\end{equation}
	\begin{equation}
		\mu_{\rm O_2}^{\rm s}=\mu_{\rm O_2}^\circ(T_{\rm s})+RT_{\rm s}\ln\left(\frac{f_{\rm O_2}^{\rm s}}{f^\circ}\right),
	\end{equation}
	where $\Delta_{\rm r} G_{\rm s}$ is the reaction Gibbs energy at the surface, $\mu_{i}^{\rm atm,s}$ and $f_{i}^{\rm atm,s}$ are the chemical potential and fugacity of species $i$ in the atmosphere at the surface, and $\mu_{\rm O_2}^{\rm s}$ and $f_{\rm O_2}^{\rm s}$ are the oxygen chemical potential and oxygen fugacity corresponding to the redox state of the magma-ocean surface. The fugacities of H$_2$ and H$_2$O are approximated by their respective partial pressures assuming ideal-gas behavior. Chemical equilibrium is attained when $\Delta_{\rm r} G_{\rm s}=0$. The corresponding equilibrium condition is
	\begin{equation}
		\frac{f_{\rm H_{2}O}^{\rm atm,s}}{f_{\rm H_2}^{\rm atm,s}} = K_{\rm eq}(T_{\rm s}) \left( \frac{f_{\rm O_2}^{\rm s}}{f^{\circ}} \right)^{1/2}.
	\end{equation}
	Once the surface redox equilibrium is established, this equilibrium condition is maintained while simultaneously satisfying the H$_2$O solubility equilibrium and the mass balances of H, O, and Fe associated with the redox reactions.

\begin{figure*}
\gridline{\fig{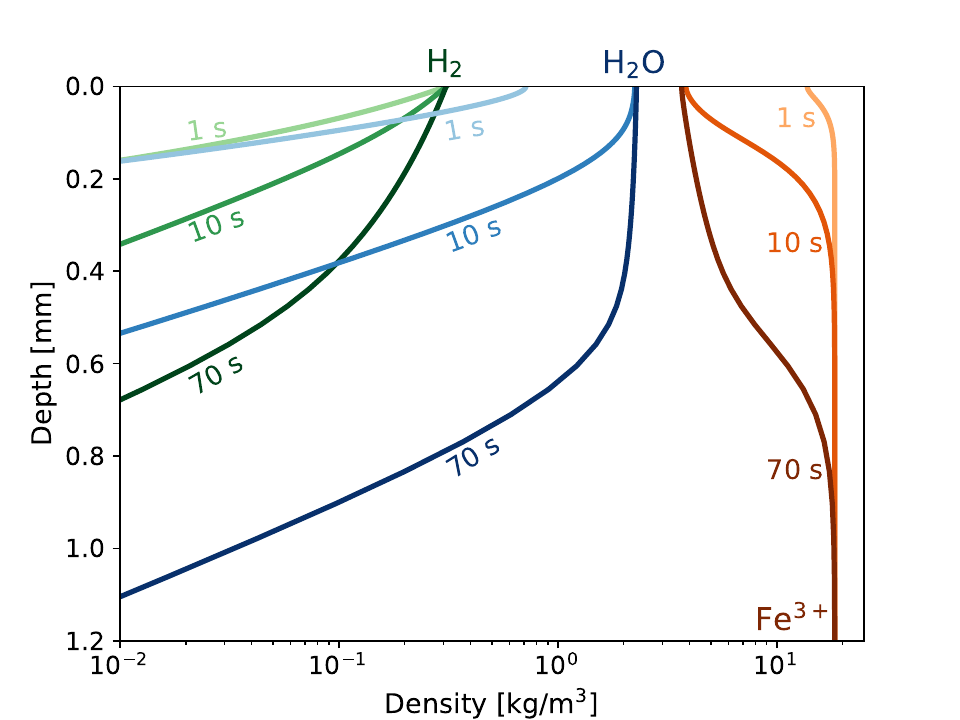}{0.45\textwidth}{(a)}
		\fig{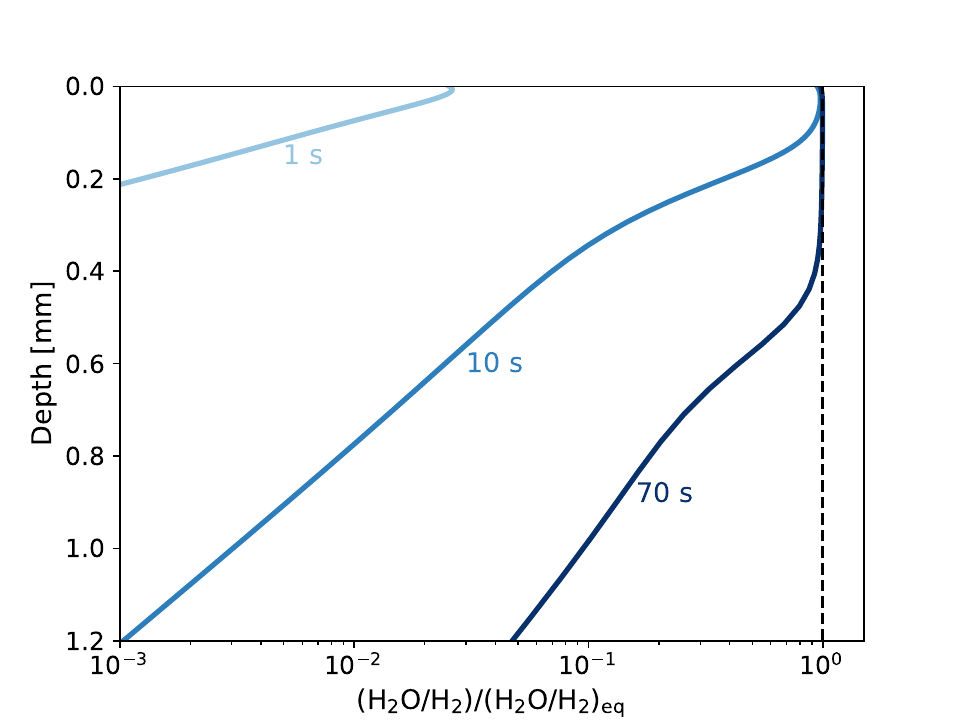}{0.45\textwidth}{(b)}
          	}
\gridline{\fig{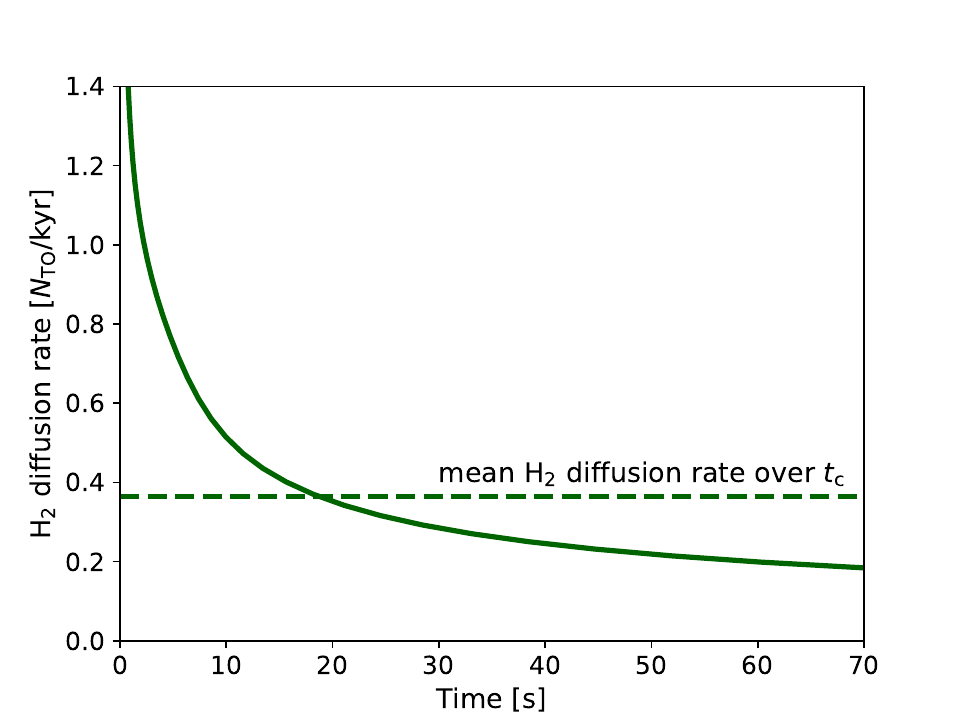}{0.45\textwidth}{(c)}
		\fig{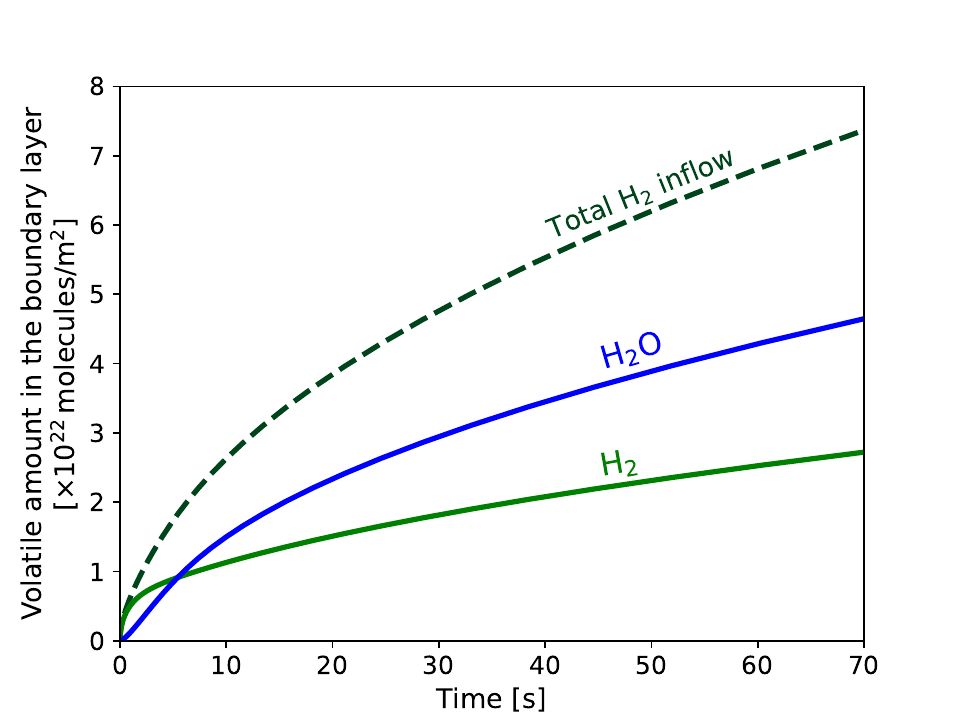}{0.45\textwidth}{(d)}
          	}
\caption{(a) Density profiles of H$_2$ (green), H$_2$O (blue), and Fe$^{3+}$ (orange) and (b) the H$_2$O/H$_2$ density ratio normalized by its equilibrium value at each depth in the boundary layer at selected times during a boundary layer overturn time of 70 s. The full boundary layer thickness calculated from Equation (34) is 19 mm, but only the upper part of the boundary layer, where H$_2$ diffusion occurs, is shown for clarity. Color gradients indicate temporal evolution. The H$_2$ surface pressure is 300 bar, the surface temperature is 2400 K, and the initial Fe$^{3+}$/Fe$^{2+}$ ratio is 0.08. The overturn time of 70 s corresponds to a surface temperature of 2400 K, a magma ocean bottom pressure of 50 GPa, and an internal heat flux of 400 W/m$^{2}$, representative of the early stage of magma ocean evolution shown in Figures~\ref{fig:6} and \ref{fig:7}. (c) Corresponding temporal evolution of the H$_2$ diffusion rate normalized by the hydrogen molecular number in the present-day terrestrial seawater ($N_{\rm TO}$). The dashed line represents the mean H$_2$ diffusion rate over the overturn time, as defined by Equation (41). (d) Corresponding temporal evolution of the amounts of H$_2$ (green) and H$_2$O (blue) in the boundary layer. The dashed line represents the cumulative H$_2$ inflow into the boundary layer.}
\label{fig:4}
\end{figure*}

\begin{figure}[htbp]
	\centering
	\includegraphics[width=\columnwidth]{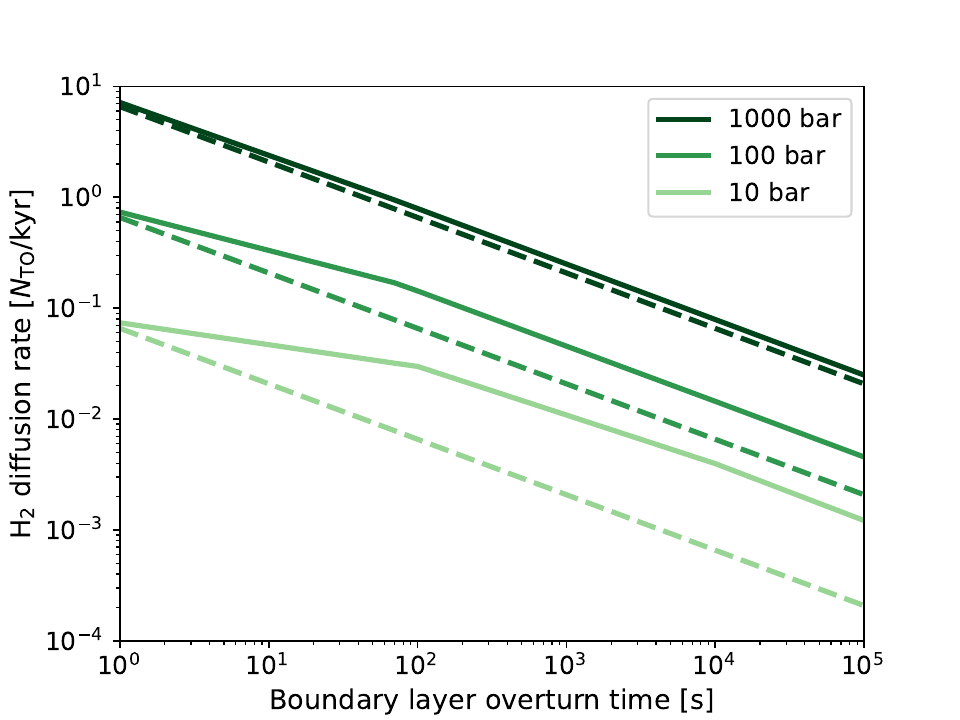}
	\caption{Mean H$_2$ diffusion rate normalized by the hydrogen molecular number in the present-day terrestrial seawater ($N_{\rm TO}$), over the boundary layer overturn time. The solid lines represent the calculation results and the dashed lines represent the analytical solution without chemical reactions described by Equation (42). The surface temperature is 2400 K and the initial Fe$^{3+}$/Fe$^{2+}$ ratio is 0.08. Colors indicate different surface H$_2$ partial pressures.}
	\label{fig:5}
\end{figure}

\begin{figure*}[htbp]
	\centering
	\includegraphics[width=0.8\textwidth]{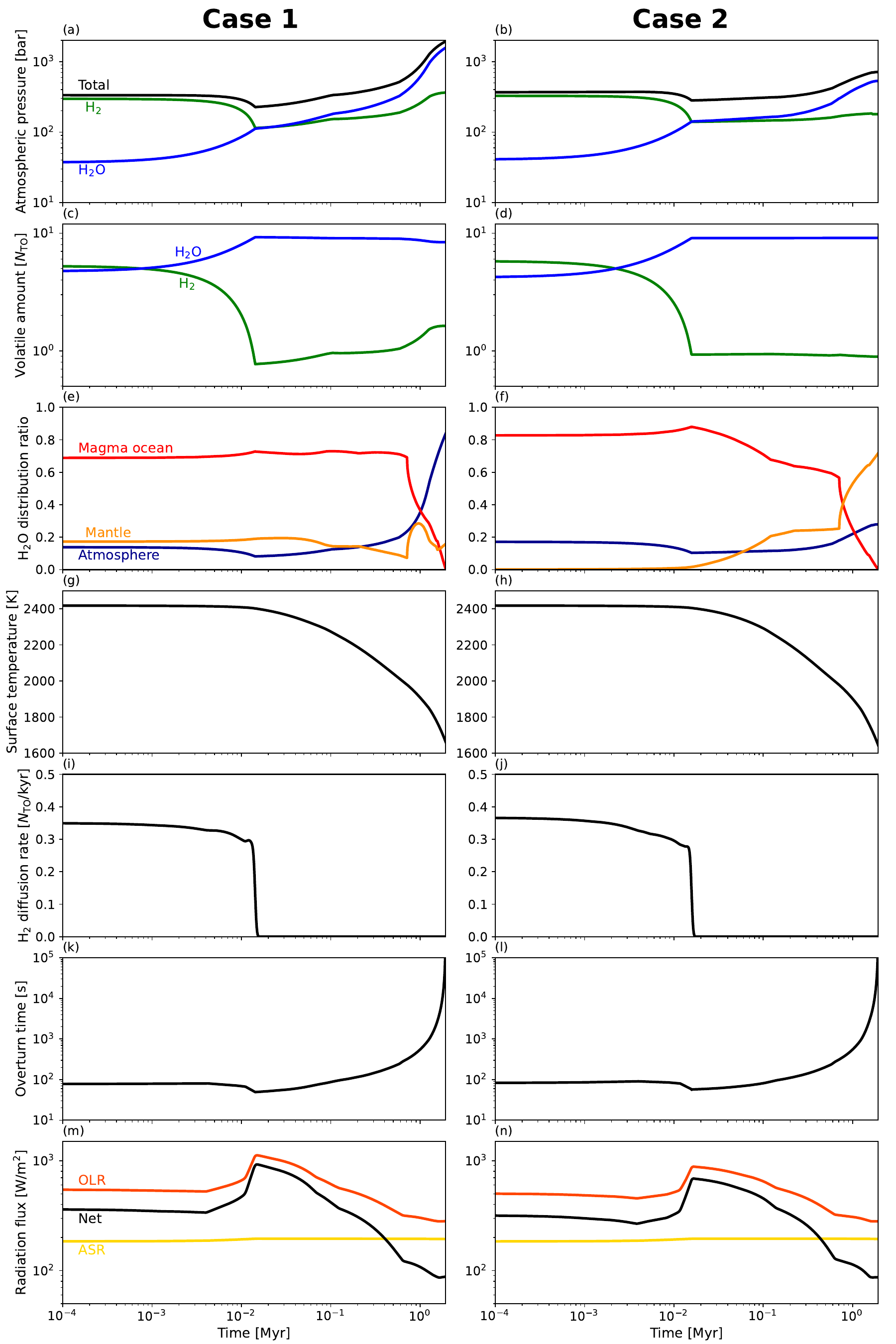}
	\caption{Example of the coupled evolution of a magma ocean and proto-atmosphere for an initial H$_2$+H$_2$O amount of $10N_{\rm TO}$ and an initial magma ocean bottom pressure of 50 GPa, focusing on the behavior of volatiles and the boundary layer. The left and right columns show the results of Case 1 and Case 2, respectively. (a, b) Atmospheric pressure of the total (black), H$_2$ (green), and H$_2$O (blue). (c, d) Total amounts of H$_2$ (green) and H$_2$O (blue) normalized by the number of hydrogen molecules in the present-day terrestrial seawater ($N_{\rm TO}$). (e, f) H$_2$O distribution ratio in the atmosphere (blue), magma ocean (red), and mantle (orange). (g, h) Surface temperature. (i, j) H$_2$ diffusion rate into the magma ocean, normalized by $N_{\rm TO}$. (k, l) Overturn time of the boundary layer. (m, n) Net outgoing radiation flux (black), outgoing longwave planetary radiation flux (red), and absorbed solar radiation flux (yellow).}
	\label{fig:6}
\end{figure*}

\begin{figure*}[htbp]
	\centering
	\includegraphics[width=0.8\textwidth]{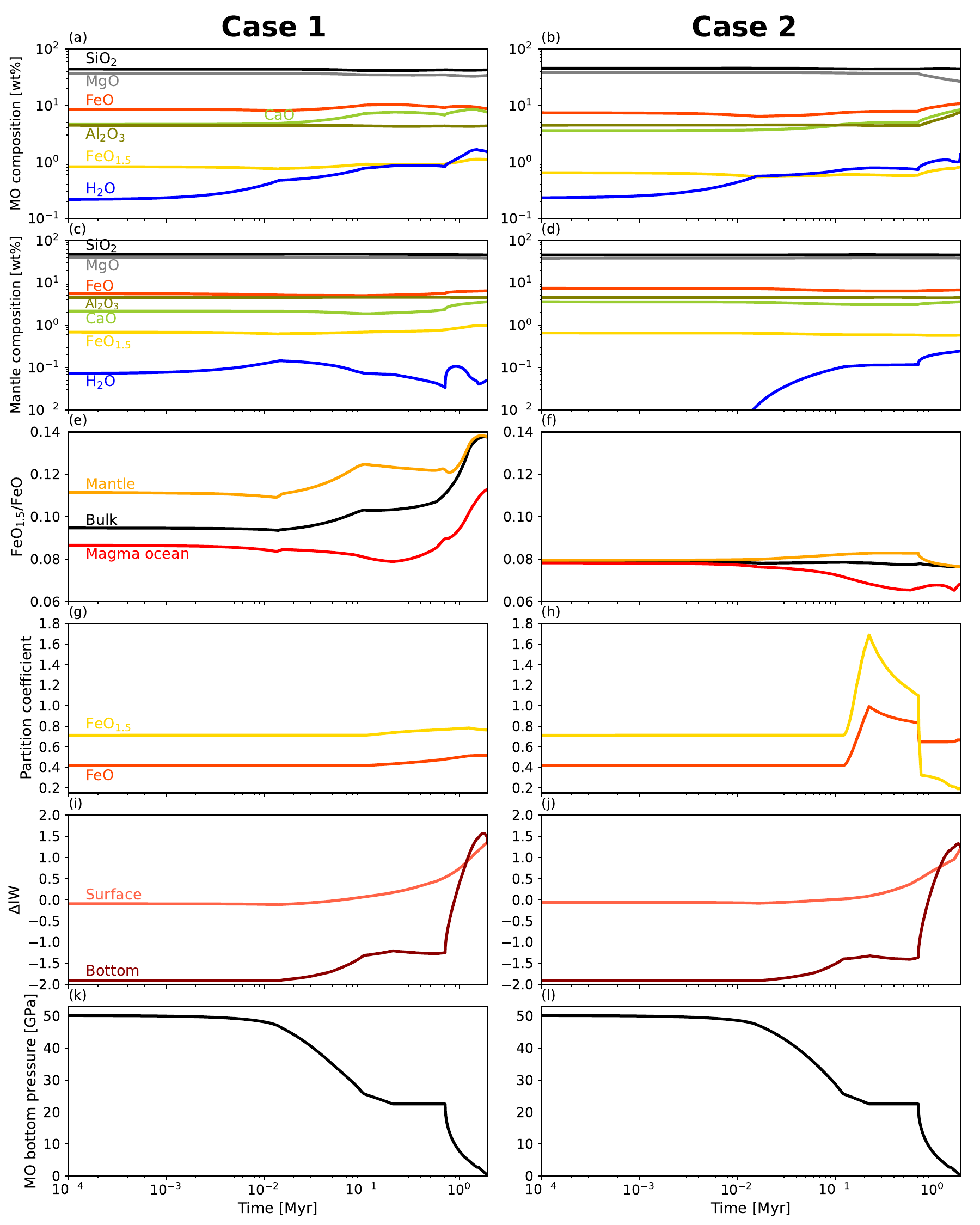}
	\caption{Example of the coupled evolution of a magma ocean and proto-atmosphere for an initial H$_2$+H$_2$O amount of $10N_{\rm TO}$ and an initial magma ocean bottom pressure of 50 GPa, focusing on the behavior of oxide components and redox state of the magma ocean. The left and right columns show the results of Case 1 and Case 2, respectively. (a, b) Magma ocean composition. (c, d) Mantle composition. (e, f) Molar FeO$_{1.5}$/FeO ratio in the bulk interior (black), magma ocean (red), and mantle (orange). (g, h) Partition coefficient of FeO (red) and FeO$_{1.5}$ (yellow). (i, j) Oxygen fugacity relative to the IW buffer at the surface (pink) and the magma ocean bottom (brown). (k, l) Pressure at the magma ocean bottom.}
	\label{fig:7}
\end{figure*}

\begin{figure*}[htbp]
	\centering
	\includegraphics[width=0.75\textwidth]{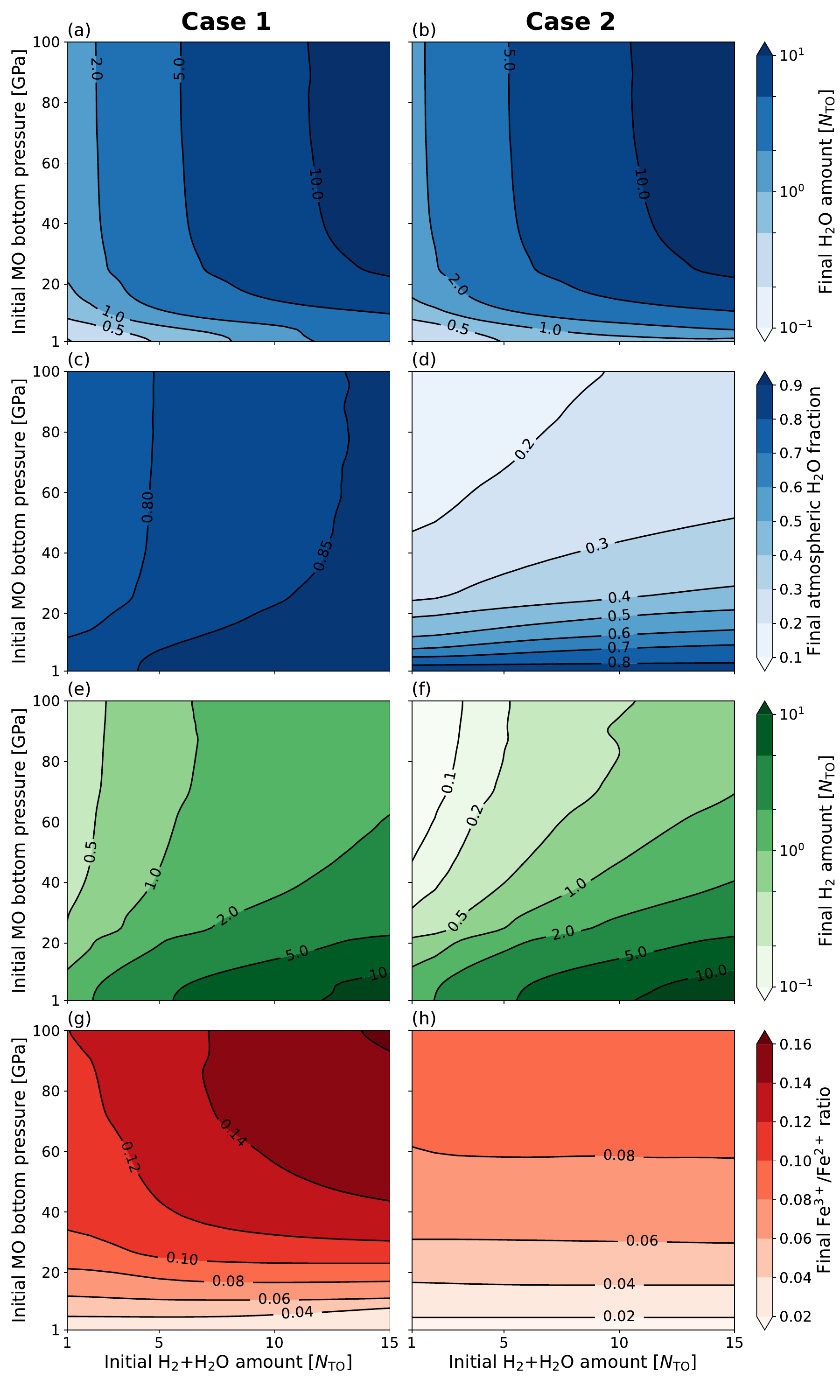}
	\caption{System state just after the magma ocean solidification is finished as a function of the initial H$_2$+H$_2$O amount, normalized by the hydrogen molecular number in the present-day terrestrial seawater ($N_{\rm TO}$), and the initial bottom pressure of the magma ocean. The left and right columns show the results of Case 1 and Case 2, respectively. (a, b) Total H$_2$O amount normalized by $N_{\rm TO}$. (c, d) Fraction of the atmospheric H$_2$O amount to the total H$_2$O amount. (e, f) Total H$_2$ amount normalized by $N_{\rm TO}$. (g, h) Fe$^{3+}$/Fe$^{2+}$ ratio in the whole mantle.}
	\label{fig:8}
\end{figure*}

\section{Results} \label{sec:result}
\subsection{Diffusive transport and H$_2$O production in the magma ocean boundary layer}
	In this subsection, we present the results of diffusive transport and H$_2$O production in the boundary layer at the magma ocean--atmosphere interface before describing the long-term evolutionary calculations. The boundary layer overturn time, given by Equation (36), is a key factor controlling the H$_2$ diffusion rate. As shown in Figures~\ref{fig:3}, the overturn time decreases with increasing surface temperature because of the lower melt viscosity and also decreases with increasing internal heat flux. During the magma-ocean stage, the surface temperature can exceed 2000 K when a deep magma ocean extending into the lower mantle region is present, while the internal heat flux can exceed $\sim10^{2}\,\mathrm{W/m^{2}}$, allowing the overturn time to become comparable to or shorter than $\sim10^{2}$ s.

	Figures~\ref{fig:4} show the temporal evolution of the density profiles of H$_2$, H$_2$O, and Fe$^{3+}$, the H$_2$ diffusion rate and cumulative inflow, and the H$_2$O production amount in the boundary layer for an overturn time of 70 s and an atmospheric H$_2$ pressure of 300 bar. The overturn time of 70 s corresponds to a surface temperature of 2400 K, a magma ocean bottom pressure of 50 GPa, and an internal heat flux of 400 W/m$^{2}$, representative of the early stage of magma ocean evolution shown in Figures~\ref{fig:6} and \ref{fig:7}. H$_2$ diffuses into the boundary layer according to the diffusion coefficient given by Equation (17), while simultaneously being consumed by reaction with Fe$^{3+}$ to produce H$_2$O and Fe$^{2+}$ (Figure~\ref{fig:4}(a)). As the reaction proceeds, the H$_2$O density increases and the H$_2$O/H$_2$ ratio progressively approaches its equilibrium value from the surface downward (Figure~\ref{fig:4}(b)).

	The H$_2$ diffusion rate at the surface decreases with time as the H$_2$ density gradient becomes shallower. Here, the mean H$_2$ diffusion rate integrated over one overturn time shown in Figure~\ref{fig:4}(c) is defined as
	\begin{equation}
		\dot{N}_{\rm H_{2}}=4\pi R_{\rm p}^{2} \bar{F}_{\rm H_{2}},
	\end{equation}
	where $N_{\rm H_{2}}$ is the total number of atmospheric H$_2$, $\dot{N}_{\rm H_{2}}$ is the mean H$_2$ diffusion rate integrated over the overturn time, and $R_{\rm p}$ is the planetary radius. Although the H$_2$ diffusion distance is much smaller than the full boundary-layer thickness of 19 mm in this case, reflecting the lower diffusivity of H$_2$ relative to thermal diffusivity, the diffusion rate remains higher than $\sim 0.1N_{\rm TO}\mathrm{/kyr}$, with a mean value of $0.35N_{\rm TO}\mathrm{/kyr}$. This high diffusion rate results from the short residence time of the boundary layer, which maintains a steep near-surface H$_2$ density gradient. Thus, despite the low diffusivity of H$_2$ in the melt, efficient H$_2$ transport into the magma ocean and subsequent H$_2$O production can be achieved through frequent renewal of the boundary layer associated with low melt viscosity and high internal heat flux during the magma-ocean stage. Although conversion of the transported H$_2$ to H$_2$O proceeds within the boundary layer, some H$_2$ remains at the time of overturn (Figure~\ref{fig:4}(d)). This residual H$_2$ is assumed to be transported into and mixed with the oxidized bulk magma ocean, where it is eventually converted to H$_2$O.
		
	The mean diffusion rate as a function of overturn time and H$_2$ partial pressure is shown in Figure~\ref{fig:5}. The mean diffusion rate increases with decreasing overturn time and increasing H$_2$ partial pressure. For H$_2$ pressures exceeding 100 bar and overturn times shorter than $\sim$10$^{2}$ s, the mean diffusion rate can exceed $\sim 0.1N_{\rm TO}/\mathrm{kyr}$, mainly owing to frequent renewal of the boundary layer described above. The dashed lines in Figure~\ref{fig:5} represent the analytical solution without chemical reactions:
	\begin{equation}
		\dot{N}_{\rm H_{2}}^{\rm analytical}=4\pi R_{\rm p}^{2} \times 2 n_{\rm H_{2}}^{0} \sqrt{\frac{D_{\rm H_{2}}^{0}}{\pi t_{\rm c}}},
	\end{equation}
	where $n_{\rm H_{2}}^{0}$ and $D_{\rm H_{2}}^{0}$ are the number density and diffusion coefficient of H$_2$ at the upper boundary corresponding to the surface. The numerical results asymptotically approach the analytical solution in the limit where the overturn time is shorter than the chemical reaction timescale of $\sim 1$ s. At longer overturn times, the numerical diffusion rate exceeds the analytical estimate because consumption of H$_2$ by reaction with Fe$^{3+}$ maintains a steeper H$_2$ density gradient and enhances continued diffusion into the magma ocean.

\subsection{Coupled evolution of proto-atmosphere and magma ocean}
	In this section, we investigate the coupled evolution of an H$_2$-rich proto-atmosphere and an oxidized magma ocean, building on the results of the diffusion of atmospheric H$_2$ into the magma ocean described in Section 3.1, and accounting for redox reactions in the magma ocean, material partitioning among coexisting phases, and planetary cooling via outgoing thermal radiation. Figure~\ref{fig:6} and \ref{fig:7} show examples of the coupled evolution, assuming an initial H$_2$+H$_2$O amount of 10$N_{\rm TO}$ and an initial magma ocean bottom pressure of 50 GPa, where $N_{\rm TO}$ is the number of hydrogen molecules equivalent to that in the present-day Earth's seawater. Here, the initial H$_2$+H$_2$O amount is motivated by estimates of the present-day BSE water inventory, while the initial magma ocean bottom pressure lies within the pressure range of metal-silicate equilibration inferred from siderophile-element partitioning, as discussed in detail in Section 4.3. Figure~\ref{fig:6} focuses on the behavior of volatiles and the boundary layer, whereas Figure~\ref{fig:7} focuses on the behavior of oxide components and the redox state of the magma ocean.
	
	Firstly, we focus on the result of Case 1, where the equilibrium distribution of H$_2$O and oxide components between the bulk magma ocean and solidified mantle is assumed (left columns of Figures~\ref{fig:6} and \ref{fig:7}). During the early stage, the diffusion of atmospheric H$_2$ into the magma ocean is efficient, with a diffusion rate of $\sim 0.35N_{\rm TO}/\mathrm{kyr}$ (Figure~\ref{fig:6}(i)), due to the short overturn time of the boundary layer (Figure~\ref{fig:6}(k)). This leads to the efficient production of H$_2$O through the redox reaction between H$_2$ and FeO$_{1.5}$ (Figure~\ref{fig:6}(a) and (c)). The relatively high oxygen fugacity at the surface, which highly depends on pressure (Figure~\ref{fig:7}(i)), further promotes this reaction. Although FeO$_{1.5}$ is consumed by the reaction with H$_2$, its concentration and the oxygen fugacity in the magma ocean remain nearly constant during this phase due to the continuous disproportionation of Fe$^{2+}$ at the base of the magma ocean (Figure~\ref{fig:7}(a), (c), and (e)). The FeO$_{1.5}$/FeO ratio in the solid mantle is larger than that in the magma ocean (Figure~\ref{fig:7}(e)), reflecting the higher partition coefficient of FeO$_{1.5}$ (Figure~\ref{fig:7}(g)) and the large fraction of bridgmanite in the bulk mantle, which preferentially incorporates FeO$_{1.5}$ \citep{Schaefer2024}.
	
	The chemical equilibrium between the atmosphere and magma ocean at the surface is achieved when the atmospheric H$_2$O/H$_2$ ratio reaches $\sim 1$ at $\sim 0.01$ Myr. Afterward, the magma ocean cools and solidifies through the planetary radiation (Figure~\ref{fig:6}(g) and (m); Figure~\ref{fig:7}(k)). The cooling rate is limited by the blanketing effect of the proto-atmosphere due to H$_2$O absorption, H$_2$-H$_2$ collision-induced absorption, and Rayleigh scattering \citep{Yoshida2025}. As solidification progresses, H$_2$O degasses from the magma ocean into the atmosphere (Figure~\ref{fig:6}(a) and (e)). Changes in the atmospheric H$_2$O/H$_2$ ratio caused by degassing drive redox reactions between H$_2$O and FeO to maintain chemical equilibrium at the surface, producing additional H$_2$ and FeO$_{1.5}$ (Figure~\ref{fig:6}(c); Figure~\ref{fig:7}(a), (c), and (e)). In other words, in Case 1, the atmosphere acts as an oxidizing agent for the solidifying magma ocean, buffering its solidification-driven reduction. The solidification of the magma ocean completes at $\sim 2$ Myr, by which time over 80\% of H$_2$O has been degassed into the atmosphere (Figure~\ref{fig:6}(e)), leaving approximately $1N_{\rm TO}$ of atmospheric H$_2$ (Figure~\ref{fig:6}(c)).
	
	In Case 2, the diffusion of atmospheric H$_2$ into the magma ocean, subsequent production of H$_2$O, and the cooling and solidification of the magma ocean through planetary radiation follow a similar pattern. The key differences lie in the partitioning of H$_2$O between the interior and the atmosphere. In Case 2, H$_2$O-rich melts within the rheological transition region are trapped inside the solidifying mantle as described by Equation (12), resulting in greater retention of H$_2$O in the interior and less efficient degassing to the atmosphere (Figure~\ref{fig:6}(f)). On the other hand, in Case 1, the low partition coefficient of H$_2$O leads to its concentration in the magma ocean and more extensive degassing (Figure~\ref{fig:6}(e)). Because H$_2$O degassing is less efficient in Case 2, the redox reaction between H$_2$O and FeO that produces H$_2$ and FeO$_{1.5}$ at the surface becomes much less active after chemical equilibrium is reached at $\sim$0.01 Myr, as shown in the smaller changes in the amounts of H$_2$ and FeO$_{1.5}$ (Figure~\ref{fig:6}(d) and Figure~\ref{fig:7}(f)).
	
	The change in the partition coefficient of oxide minerals is also different between the two cases (Figure~\ref{fig:7}(g) and (h)). In Case 2, FeO$_{1.5}$ tends to be concentrated in the magma ocean relative to FeO as the solidification proceeds in the upper mantle region after $\sim 0.7$ Myr, reflecting its lower partition coefficient at pressures below $\sim 20$ GPa, due to its incompatibility with olivine and pyroxene (Figure~\ref{fig:7}(h)). On the other hand, its actual concentration in the magma ocean changes little during this stage (Figure~\ref{fig:7}(f)) because of continuous consumption through the redox reaction with H$_2$. Thus, in contrast to Case 1, the atmosphere acts as a reducing agent for the magma ocean, buffering its solidification-driven oxidation. The amount of H$_2$ varies only slightly during this phase (Figure~\ref{fig:6}(d)), owing to the limited amount of FeO$_{1.5}$, as discussed in detail in Section 4.2.
	
	Despite the substantial differences in material partitioning among the atmosphere, magma ocean, and solid mantle between Cases 1 and 2, the atmospheric composition and magma ocean oxygen fugacity evolve in broadly similar directions (Figure~\ref{fig:6}(a) and (b); Figure~\ref{fig:7}(i) and (j)). This similarity arises because redox exchange with the atmosphere buffers the redox state of the solidifying magma ocean. The buffering effect is described in more detail in Section 4.1.
	
	Figure~\ref{fig:8} summarizes the system state just after the magma ocean solidification is finished as a function of the initial H$_2$+H$_2$O amount and initial magma ocean bottom pressure. In both Case 1 and Case 2, the final H$_2$O amount increases with the initial bottom pressure of the magma ocean, reflecting the more oxidized surface conditions promoted by Fe$^{2+}$ disproportionation under higher pressures (Figure~\ref{fig:8}(a) and (b)). In Case 1, the final atmospheric H$_2$O fraction exceeds $\sim 80\%$, almost independently of the initial conditions (Figure~\ref{fig:8}(c)), while in Case 2, it tends to be lower especially at higher initial pressures (Figure~\ref{fig:8}(d)), due to the H$_2$O-rich melt trapping. In contrast to the result of the final H$_2$O amount, the remaining amount of H$_2$ decreases with increasing the initial bottom pressure of the magma ocean in both cases (Figure~\ref{fig:8}(e) and (f)). The final H$_2$ amount in Case 1 is slightly larger than that in Case 2 overall, owing to its production and degassing associated with H$_2$O degassing during solidification. The final Fe$^{3+}$/Fe$^{2+}$ ratio in the mantle increases with the initial magma ocean bottom pressure (Figure~\ref{fig:8}(g) and (h)), and Case 1 generally yields a higher ratio than Case 2 because of the production of FeO$_{1.5}$ with H$_2$ during solidification.

\section{Discussion} \label{sec:discussion}
\subsection{Comparison with calculations treating the mantle and magma ocean as effectively infinite oxygen reservoirs}
\begin{figure*}[htbp]
	\centering
	\includegraphics[width=0.8\textwidth]{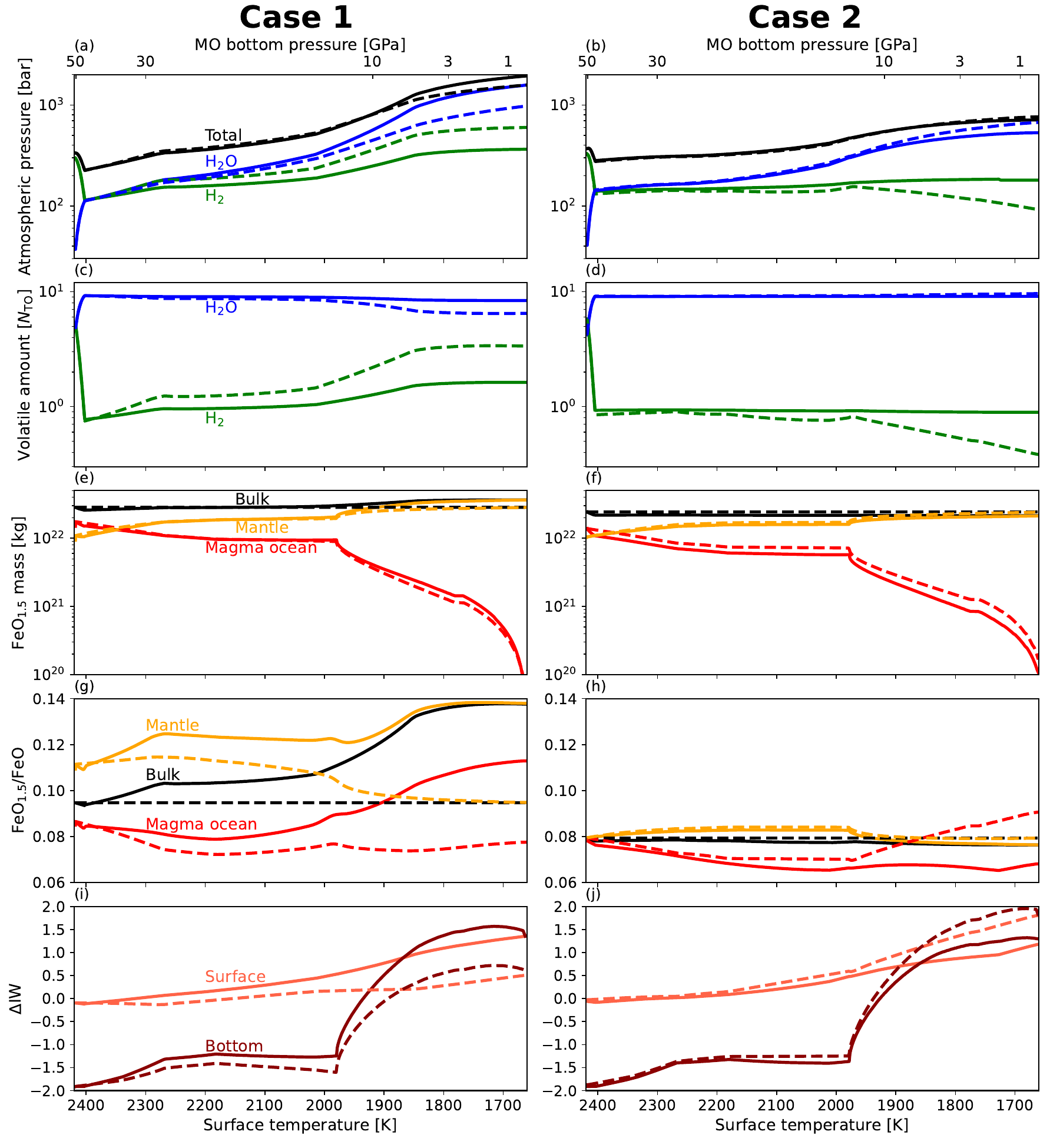}
	\caption{Comparison between the evolution calculated in this study, which includes the oxygen mass balance associated with redox reactions between the atmosphere and magma ocean (solid lines), and a calculation that neglects the atmospheric influence on the oxygen fugacity of the magma ocean, in which the atmospheric composition is determined as a dependent variable of the magma-ocean oxygen fugacity (dashed lines). The initial H$_2$+H$_2$O amount is $10N_{\rm TO}$ and the initial magma ocean bottom pressure is 50 GPa, as in Figures~\ref{fig:6} and \ref{fig:7}. The left and right columns show the results of Case 1 and Case 2, respectively. The lower horizontal axis is the surface temperature, and the upper horizontal axis is the magma ocean bottom pressure. (a, b) Atmospheric pressure of the total (black), H$_2$ (green), and H$_2$O (blue). (c, d) Total amounts of H$_2$ (green) and H$_2$O (blue). (e, f) FeO$_{1.5}$ mass in the bulk interior (black), magma ocean (red), and mantle (orange). (g, h) Molar FeO$_{1.5}$/FeO ratio in the bulk interior (black), magma ocean (red), and mantle (orange). (i, j) Oxygen fugacity relative to the IW buffer at the surface (pink) and the magma ocean bottom (brown).}
	\label{fig:9}
\end{figure*}

	Previous magma ocean--atmosphere models for Earth have often assumed that the mantle and magma ocean act as effectively infinite oxygen reservoirs, thereby neglecting the oxygen mass balance associated with redox reactions between the atmosphere and magma ocean \citep[][]{Hirschmann2012,Pahlevan2019,Katyal2020,Bower2022,Maurice2024,Nicholls2024}. Under this assumption, the redox evolution of the magma ocean is computed independently of the atmosphere, and the atmospheric composition is then determined as a dependent variable from the oxygen fugacity of the magma ocean. Thus, without this coupling, the atmospheric composition inherits the dependence of the magma ocean redox state on solidification-driven changes in magma-ocean composition, which are associated with crystal--melt partitioning and material exchange between the magma ocean and solid mantle \citep[e.g.,][]{Schaefer2024}. In this section, we compare our results obtained with this oxygen mass balance to those calculated under this conventional assumption, which neglects atmospheric influence on the oxygen fugacity of the magma ocean, as shown in Figure~\ref{fig:9}. Here, the initial H$_2$+H$_2$O amount is $10N_{\rm TO}$ and the initial magma ocean bottom pressure is 50 GPa, as in Figures~\ref{fig:6} and \ref{fig:7}. The horizontal axis is expressed in terms of surface temperature and magma ocean basal pressure, rather than time.
	
	In Case 1, the magma ocean tends to become more reducing than the atmosphere during the solidification through the preferential partitioning of FeO$_{1.5}$ into the mantle. Consequently, the calculated H$_2$O/H$_2$ ratio, FeO$_{1.5}$ mass, FeO$_{1.5}$/FeO ratio, and oxygen fugacity are all lower when the oxygen mass balance associated with redox reactions between the atmosphere and magma ocean is neglected (left panels of Figure~\ref{fig:9}). This difference arises because the calculation ignoring this oxygen mass balance neglects the production of FeO$_{1.5}$ through the redox reaction between H$_2$O and FeO, when H$_2$O degassing associated with solidification proceeds. The comparison demonstrates that, in our oxygen-mass-balanced calculation in Case 1, the atmosphere acts to oxidize the magma ocean to maintain chemical equilibrium during solidification. In other words, redox exchange with the atmosphere buffers the solidification-driven reduction of the magma ocean.
	
	In Case 2, by contrast, the magma ocean becomes more oxidized as FeO$_{1.5}$ accumulates when the upper-mantle region with a pressure lower than $\sim 23$ GPa solidifies. As a result, the estimated H$_2$O/H$_2$ ratio, FeO$_{1.5}$ mass, FeO$_{1.5}$/FeO ratio, and oxygen fugacity are higher in the calculation neglecting the oxygen mass balance associated with redox reactions between the atmosphere and magma ocean (right panels of Figure~\ref{fig:9}). Thus, in Case 2, the atmosphere acts to reduce the magma ocean to maintain the chemical equilibrium in our oxygen-mass-balanced calculation. In this case, redox exchange with the atmosphere buffers the solidification-driven oxidation of the magma ocean. On the other hand, the amount of H$_2$ changes little during this phase (Figure~\ref{fig:9}(d)) because of the limited availability of FeO$_{1.5}$ in the magma ocean (Figure~\ref{fig:9}(f)).
	
	In summary, our results show that the atmosphere can buffer the redox state of the magma ocean during solidification, acting as either an oxidizing or reducing buffer that prevents the atmosphere-magma ocean system from undergoing extreme redox divergence. The important point is that this buffering effect reduces the sensitivity of the redox evolution of the atmosphere--magma ocean system to solidification-driven changes in magma ocean composition caused by different efficiencies of material exchange between the magma ocean and solid mantle. As shown by the comparison between the left and right columns in Figure~\ref{fig:9}, when the oxygen mass balance associated with redox reactions between the atmosphere and magma ocean is included, redox exchange between the atmosphere and magma ocean buffers changes in the magma-ocean redox state during solidification, and both cases evolve toward broadly similar atmospheric compositions and magma ocean oxygen fugacities. By contrast, when this oxygen mass balance is neglected, the atmospheric composition and magma ocean oxygen fugacity differ substantially between Cases 1 and 2, reflecting their different assumptions about material exchange between the magma ocean and solid mantle and the resulting differences in solidification-driven changes in magma-ocean composition, consistent with the sensitivity of magma ocean redox evolution to solidification processes shown by \citet{Schaefer2024}. These differences in the redox evolution of the atmosphere-magma ocean system can substantially affect estimates of the atmosphere and surface redox conditions after the solidification of the magma ocean.

\subsection{Effects of model uncertainties and simplifications}
	In this study, we assume that metallic iron droplets formed via the disproportionation of Fe$^{2+}$ rapidly sink into the core, thereby enabling efficient oxidation of the magma ocean through an increase in the Fe$^{3+}$/Fe$^{2+}$ ratio. This assumption is supported by estimates of droplet size and settling velocity. According to Stokes' law, a stable droplet with a radius of $\sim 1$ cm in a silicate melt (assuming typical magma ocean viscosities) would settle at a velocity of $\sim 0.5\,\mathrm{m/s}$ \citep{Rubie2003,Rubie2007,Lichtenberg2021b}, which is significantly faster than the diffusion velocity of atmospheric H$_2$. However, metal droplets could be entrained by vigorous convection in the magma ocean if the coagulation is inefficient \citep{Rubie2007,Lichtenberg2021b}. In such cases, the oxygen fugacity near the surface could remain buffered at the IW buffer, suppressing the oxidation of H$_2$ and thereby limiting H$_2$O production.
	
	Although we apply the formulation of Fe$^{3+}$/Fe$^{2+}$ ratio as a function of pressure and temperature given by \citet{Zhang2024}, the estimates by high-pressure experiments and first‐principle calculations show highly divergent Fe$^{3+}$/Fe$^{2+}$ values. \citet{Armstrong2019} and \citet{Kuwahara2023} reported higher Fe$^{3+}$/Fe$^{2+}$ ratios than measured by \citet{Zhang2024} under similar conditions, whereas the first-principle calculations by \citet{Deng2020} predicted a more moderate Fe$^{3+}$/Fe$^{2+}$ ratio. If the actual Fe$^{3+}$/Fe$^{2+}$ ratio is higher than assumed here, the surface oxygen fugacity would also be elevated, leading to more efficient production of H$_2$O through redox reactions with atmospheric H$_2$.
	
	To account for uncertainties in the efficiency of material exchange between the magma ocean and solidifying mantle, we have considered two endmember cases (Case 1 and Case 2). Achieving complete equilibrium partitioning between the bulk magma ocean and the mantle, as assumed in Case 1, is likely unrealistic given the slow rates of melt percolation and solid diffusion \citep[e.g.,][]{Solomatov2007}. Conversely, melts trapped in the mantle may be transported upward by convective motion during solidification, although this upward transport is neglected in Case 2. Therefore, the actual efficiency of material exchange between the magma ocean and the mantle probably lies between these two endmember assumptions.
	
	Large uncertainties also remain in the partition coefficients of oxide components, which affect their distribution between the magma ocean and mantle \citep{Schaefer2024}. Nevertheless, changes in the partitioning of oxide components are expected to have only a minor impact on the atmospheric composition evolution during solidification, because concentrations or depletions of minerals with solidification contribute little additional oxidizing or reducing capacity at this stage (see Section 4.1). Variations in the critical melt fraction defining the rheological transition can further affect the degree of partitioning between the magma ocean and mantle; in particular, the amount of H$_2$O trapped in the mantle in Case 2 depends almost linearly on this critical melt fraction.
	
	To estimate the diffusive transport in the boundary layer at the interface between the magma ocean and atmosphere, we adopt the experimental data of the diffusion coefficients in silicate melts \citep{Zhang2010a,Zhang2010b}. However, these data are limited to temperatures below $\sim 1700\,\mathrm{K}$, and we conservatively apply the values at 1700 K for higher temperatures. Since diffusion coefficients generally increase with temperature, this likely underestimates the true H$_2$ diffusion rate under hotter conditions. Similarly, the rate constant for the redox reaction between H$_2$ and Fe$^{3+}$ is based on experimental data for solid iron oxides from \citet{Barde2016}, as corresponding high-temperature data in silicate melts are lacking. The reaction rate is also expected to be higher in molten conditions, which would further enhance the efficiency of H$_2$ consumption, particularly in regimes where the boundary layer overturn time exceeds the chemical reaction timescale (see Figure~\ref{fig:4}).
	
	We estimate the boundary layer overturn time using the transient boundary layer theory \citep{Turcotte2002}, in which the overturn time is determined from the condition for buoyancy-driven convective instability of the cold, dense boundary layer formed by surface cooling. By analogy with gas exchange at the present-day air-sea interface, other complex multi-scale processes can renew or disrupt the boundary layer. The gas transfer velocity across the air-sea interface is estimated to be strongly enhanced by wind-driven turbulence and surface renewal \citep[e.g.,][]{Wanninkhof1992,Wanninkhof2009}. Wave breaking and the associated bubbles and droplets can further enhance gas exchange by disrupting and renewing the interfacial boundary layer and by providing additional gas-transfer pathways \citep[e.g.,][]{Woolf1997,Deike2022}. If similar surface-renewal processes operate at the magma-ocean surface, they would tend to enhance the effective H$_2$ transport into the magma ocean compared to the estimates from the purely diffusive framework in this study, further supporting our conclusion that H$_2$ transport into the magma ocean and subsequent H$_2$O production can proceed efficiently.
	
	Although we assume that the atmosphere is fully convective below the stratosphere following the majority of previous studies, thick proto-atmospheres can become stable against convection and develop a nearly isothermal radiative layer in high-pressure regions, due to the absorption of most stellar irradiation in the upper atmosphere \citep{Selsis2023,Nicholls2024}. Such convective shutdown can enhance outgoing thermal radiation by increasing the temperature of the upper emitting layers, as discussed in \citet{Yoshida2025}. Conversely, in H$_2$-rich atmospheres, H$_2$O condensation can suppress convection by producing a vertical gradient in the atmospheric mean molecular weight, leading to the formation of a deep radiative layer and reducing outgoing thermal radiation \citep{Leconte2017,Leconte2024}. These effects can quantitatively modify the magma-ocean solidification timescale, but the chemically coupled evolution between the magma ocean and atmosphere is not expected to be strongly affected as long as the redox reaction timescale remains shorter than the solidification timescale. At the same time, suppressed atmospheric mixing associated with convective shutdown or H$_2$O degassing could affect the supply of H$_2$ to, and the removal of H$_2$O from, the atmosphere--magma ocean interface. Although this study focuses on the magma-ocean-side boundary layer because gas transport is expected to be limited mainly by diffusion in the silicate melt, whose molecular diffusivity is orders of magnitude smaller than that in the atmosphere \citep{Olson2018,Olson2019}, the possible influence of atmospheric mixing remains an important uncertainty.
	
	Although this study does not consider chemical reactions between H$_2$ and vaporized rocky materials in the atmosphere, such reactions could produce H$_2$O after a giant impact \citep[e.g.,][]{Charnoz2023,Kurosaki2023}, with the amount of H$_2$O produced depending strongly on the composition and abundance of the vaporized rocky materials. If a large amount of oxidized rocky material is supplied to the atmosphere, for example under conditions sufficiently hot for efficient rock vaporization, and is not strongly buffered by coexisting metallic iron, such reactions could provide an additional pathway for H$_2$O production, complementary to the melt-phase reaction considered in this study. In addition, we neglect the bulk partitioning of H$_2$ into the interior. If dissolved H$_2$ is transported into regions of the magma ocean with sufficient oxidizing capacity, it could potentially enhance local H$_2$ oxidation and H$_2$O production. On the other hand, because the oxygen fugacity in our model generally decreases with depth, dissolved H$_2$ transported into deeper, more reduced regions would not necessarily be efficiently converted into H$_2$O.

\begin{figure*}[htbp]
	\centering
	\includegraphics[width=0.8\textwidth]{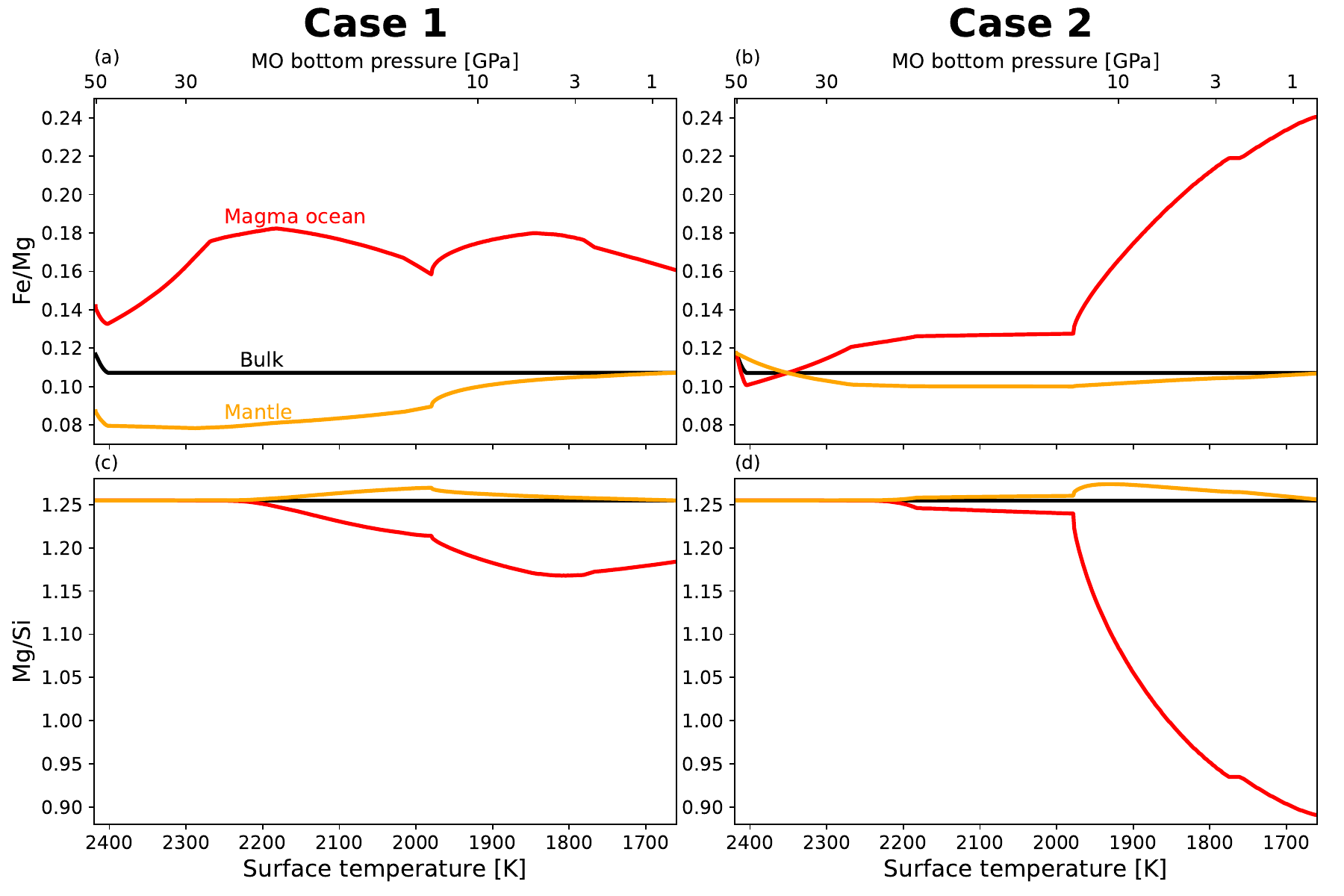}
	\caption{The compositional evolution of a magma ocean for an initial H$_2$+H$_2$O amount of $10N_{\rm TO}$ and an initial magma ocean bottom pressure of 50 GPa, as in Figures~\ref{fig:6} and \ref{fig:7}. The left and right columns show the results of Case 1 and Case 2, respectively. (a, b) Fe/Mg ratio in the bulk interior (black), magma ocean (red), and mantle (orange). (c, d) Mg/Si ratio in the bulk interior (black), magma ocean (red), and mantle (orange).}
	\label{fig:10}
\end{figure*}

\subsection{Consistency with the characteristics of the present-day Earth and Mars}
	Our results suggest that H$_2$O can be produced efficiently by the chemical interaction between an H$_2$-rich proto-atmosphere and oxidized magma ocean. According to our calculations, atmospheric H$_2$ is largely converted to H$_2$O, resulting in an H$_2$O fraction exceeding 70\% by the end of solidification when the initial bottom pressure of the magma ocean falls within the range of metal-silicate equilibration (28-53 GPa) inferred from siderophile-element partitioning \citep[e.g.,][]{Hirschmann2022} (Figure \ref{fig:8}(a), (b), (e), and (f)). Thus, if the proto-Earth acquired its atmosphere with H$_2$ inventory comparable to the current BSE's H$_2$O content with $\sim 1-10N_{\rm TO}$ \citep[e.g.,][]{Genda2016,Peslier2017,Ohtani2021} either through accreting materials or captured nebular gas, this chemical interaction could explain the present-day water budget. 
	
	Earth's building blocks could have supplied ample hydrogen. Recent cosmochemical studies suggest that most of Earth’s accreting materials closely resembled enstatite chondrites throughout the accretion phase \citep[e.g.,][]{Dauphas2017,Halliday2023}. Although enstatite chondrites are the driest among primitive meteorites, they contain hydrogen at levels exceeding Earth's seawater mass fraction \citep{Piani2020}. Combined with contributions from minor water-rich components, such as carbonaceous chondrite-like materials, and the gravitational capture of solar nebula gas, the proto-Earth could have acquired a sufficient hydrogen inventory to generate the BSE’s water via chemical reaction with an oxidized magma ocean. Although H$_2$ can escape to space, especially by hydrodynamic escape driven by intense X-ray and extreme ultraviolet irradiation from the young Sun, the effect should be negligible because of the low escape rate with $< \sim 1\,\mathrm{bar/Myr}$ due to the radiative cooling effects of C-bearing species and H$_2$O \citep{Yoshida2020,Yoshida2021,Yoshida2022,Yoshida2024b}.
	
	In this scenario, the oxidizing capacity required for H$_2$O production could also be supplied, even though enstatite-chondrite-like materials are highly reduced and FeO-poor, through FeO production associated with Si partitioning into metallic iron via the reaction SiO$_2$ + 2Fe = Si + 2FeO \citep{Javoy1995,Javoy2010}, as well as through contributions from more oxidized accreting materials. 
	
	The hydrogen isotopic composition further constrains the origin of Earth's hydrogen. The D/H ratios observed in the mantle closely match those of enstatite chondrites \citep{Piani2020}, consistent with water production from hydrogen supplied from accreting materials like enstatite chondrites. In contrast, the D/H ratio of the terrestrial seawater is higher by several tens of permil than the mantle value. This may reflect fractionation via hydrogen escape and/or the mixing with water from carbonaceous chondrite or comet-like accreting materials, which have D/H ratios comparable to or larger than the seawater value \citep[e.g.,][]{Broadley2022}. Furthermore, the D/H ratio in igneous rocks derived from the deep mantle is low with $\delta D$ from $-218$ to $-98$  permil \citep{Hallis2015}. This may imply the contribution from the captured solar nebula gas \citep{Saito2020} and/or accretion of dust grains with adsorbed H$_2$O from the protosolar nebula \citep{Hallis2015}.
	
	Our model also predicts that, during solidification, residual melts evolve toward higher Fe/Mg and lower Mg/Si due to the preferential incorporation in the solid mantle, especially in Case 2 (Figure~\ref{fig:10}). Consequently, late-stage Fe-Si-rich cumulates can become gravitationally unstable, leading to overturn that transports these dense materials into the deep mantle \citep{Elkins2003}. This process may be consistent with the slightly lower Mg/Si inferred for the present-day lower mantle compared to the upper mantle \citep[e.g.,][]{Murakami2024}. Moreover, the compositional evolution of the magma ocean can account for the Moon’s higher FeO content compared to the bulk silicate Earth. \citet{Karato2014} and \citet{Hosono2019} have proposed that a substantial portion of the ejected Moon-forming material would derive from a magma ocean if a giant solid impactor hit the proto-Earth while it was covered with the magma ocean. In that scenario, FeO-rich residual melts can explain the Moon’s elevated FeO abundance, as well as its isotopic similarity to Earth. On the other hand, the oxidation of H$_2$ to H$_2$O tends to decrease the magma ocean lifetime, which can reduce the efficiency of the magma ocean differentiation and the probability of giant impacts occurring during the magma ocean phase. The effects of the magma ocean lifetime are discussed in detail in \citet{Yoshida2025}.
	
	The calculated Fe$^{3+}$/Fe$^{2+}$ ratio of the whole mantle just after the solidification is $\sim$0.06-0.15 for initial bottom pressures of the magma ocean within the metal-silicate equilibration pressure (28-53 GPa; Figure~\ref{fig:8}(g) and (h)). This is higher than the present-day mantle value (0.02-0.06) \citep[e.g.][]{Hirschmann2022}, implying that subsequent processes reduced mantle oxidation. For example, \citet{Hirschmann2022} has proposed that the production of Cr$^{3+}$ by the reaction between Fe$^{3+}$ and Cr$^{2+}$ during crystallization can destroy at least 0.35wt\% Fe$^{3+}$, corresponding to a decrease in the Fe$^{3+}$/Fe$^{2+}$ by $\sim 0.04$. In addition, late accretion after the magma ocean solidification and core formation can also reduce the Fe$^{3+}$/Fe$^{2+}$ ratio if the accretion materials were chemically reduced like non-carbonaceous chondrites, as estimated by the isotopic analysis of highly siderophile elements \citep[e.g.,][]{Fischer2017,Halliday2023}. Assuming that the total mass of the late accretion materials was 0.5wt\% of the Earth's mass \citep{Day2016} with 30wt\% metallic iron and all the Fe was oxidized through the reaction with Fe$^{3+}$, the Fe$^{3+}$/Fe$^{2+}$ ratio is estimated to have decreased by up to $\sim 0.07$. Accounting for these post-solidification reduction processes, the modeled oxidation state after solidification can reach that of the present-day mantle. Although late accretion can reduce surface H$_2$O by converting at most $\sim 2N_{\rm TO}$ of H$_2$O into H$_2$, part of the H$_2$O can remain when the production of H$_2$O via chemical interactions between the proto-atmosphere and the magma ocean, as investigated in this study, is taken into account.
	
	Finally, the dependence of the Fe$^{3+}$/Fe$^{2+}$ ratio on magma ocean depth is qualitatively consistent with the difference in the mantle redox state between Mars and Earth. Oxygen fugacities inferred from Martian meteorites are around the IW buffer and significantly lower than Earth's mantle values \citep[e.g.,][]{Wadhwa2001}. If a magma ocean was formed on Mars, its bottom pressure should have been lower than that the core-mantle boundary pressure of 18-19 GPa \citep{Stahler2021}. Our results indicate that such lower pressures yield oxygen fugacities and Fe$^{3+}$/Fe$^{2+}$ ratios substantially below those for Earth’s magma ocean (Figure \ref{fig:8}(g) and (h)), consistent with the reduced redox state inferred for the Martian mantle.

\subsection{Atmospheric evolution and surface environment after the magma ocean solidification}
	After the magma ocean solidified and the convective heat flux declined, atmospheric water vapor should have condensed to form a surface ocean. Subsequently, H$_2$ likely remained along with C- and N-bearing species in the atmosphere. Late accretion may have generated additional H$_2$ via the chemical reduction with surface water. Assuming the composition and total mass of late accretion materials as supposed in the previous section, $\sim 2 N_{\rm TO}$ of H$_2$ can have been formed through this process. Under such H$_2$-rich conditions, CH$_4$ and NH$_3$ could have been efficiently synthesized through thermochemical reactions between H$_2$ and C- and N-bearing species during atmospheric cooling \citep{Zahnle2020,Wogan2023}. The resulting atmosphere enriched in the reduced species should have evolved through the atmospheric escape and photochemical processes. H$_2$ would have gradually escaped to space by hydrodynamic escape \citep{Yoshida2021,Yoshida2024b}. Concurrently, reduced species such as CH$_4$ and NH$_3$ would have undergone photochemical conversion into a variety of organic compounds and more oxidized products. CH$_4$ photolysis would primarily produce heavier organics due to their UV self-shielding effects, which suppress the H$_2$O photolysis and the subsequent production of oxidant radicals such as OH \citep{Yoshida2024a}. As a result, various and massive organics, including prebiotically important matters, may have accumulated on the surface ocean, which may have played a crucial role in the emergence of living organisms.

\section{Conclusion}
	In this study, we have developed a coupled atmosphere--interior evolution model that incorporates diffusive transport in the boundary layer between the magma ocean and atmosphere, redox reactions within the melt, partitioning of oxide components and H$_2$O among coexisting phases, and planetary cooling governed by radiative transfer through the proto-atmosphere, with metallic iron produced by Fe$^{2+}$ disproportionation removed into the core as it forms, to investigate how a magma ocean oxidized by Fe$^{2+}$ disproportionation affects an overlying H$_2$-rich proto-atmosphere during the post-core-formation stage. Specifically, we have quantified the atmospheric influence on the redox state of the solidifying magma ocean while accounting for material exchange between the magma ocean and solid mantle, including uncertainty in its efficiency, and constrained the timescale for chemical interaction between the atmosphere and magma ocean through diffusive transport in the boundary layer at the interface. Atmospheric H$_2$ transport into the magma ocean and subsequent H$_2$O production become more efficient with increasingly frequent boundary layer renewal driven by buoyant overturn of dense melt cooled at the surface, allowing the atmosphere and magma ocean to chemically equilibrate within $\sim10^4$ yr when the magma ocean is sufficiently deep and hot to extend into the lower mantle. Redox exchange with the atmosphere buffers the solidification-driven redox evolution of the magma ocean, preventing extreme oxidation or reduction of the coupled system. These results suggest that the coupled evolution could have contributed to early Earth's water inventory while maintaining a reducing H$_2$-rich atmosphere.\\
	
	We thank the anonymous reviewers for their constructive comments, which greatly improved the manuscript. We also thank Hideharu Kuwahara for valuable discussions on Fe$^{2+}$ disproportionation in magma oceans. T.Y. is supported by JSPS KAKENHI Grant Numbers 26K17214, 25K01062, and 22KK0044. N.T. is supported by JSPS KAKENHI grant numbers JP22H00164 and JP24K21562. T.K. is supported by JSPS KAKENHI grant numbers 25K01051 and 22H00164. K.K. is supported by JSPS KAKENHI grant numbers 24K07110, 22K21344, and 22H00164.
	
\bibliographystyle{aasjournal} 
\bibliography{references} 	
	
\appendix
\section{Details of the atmospheric structure}
	We suppose that atmospheres are composed of H$_{2}$O and H$_{2}$, which are expected to be major radiative absorption sources in proto-atmospheres \citep{Lichtenberg2021}. The atmosphere in hydrostatic equilibrium is vertically divided into 500 layers from the ground to the top of the atmosphere, which are equally spaced on a logarithmic scale of pressure. The pressure at the top of the atmosphere is fixed at 1 Pa. We treat H$_{2}$ as an ideal gas and H$_{2}$O as a non-ideal gas following the Peng-Robinson equation of state \citep{Peng1976}, in a similar manner to \citet{Abe1988} and \citet{Hamano2015}. The Peng-Robinson equation of state is described as follows:
	\begin{equation}
		P=\frac{RT}{V-b}-\frac{a(T)}{V(V+b)+b(V-b)},
	\end{equation}
	where $P$ is the pressure, $R$ is the gas constant, $T$ is the temperature, and $V$ is the molar volume \citep{Peng1976}. Here $a(T)$ and $b$ are defined by
	\begin{equation}
		a(T)=0.45724\frac{R^{2}T_{\rm c}^{2}}{P_{\rm c}}\left[1+\kappa \left(1-\sqrt{\frac{T}{T_{\rm c}}}\right)\right]^{2},
	\end{equation}
	\begin{equation}
		b=0.07780\frac{RT_{\rm c}}{P_{\rm c}},
	\end{equation}
	\begin{equation}
		\kappa = 0.37464 + 1.54226\omega_{\rm a} - 0.26992\omega_{\rm a}^{2},
	\end{equation}
	where $T_{\rm c}$ and $P_{\rm c}$ are the temperature and pressure at the critical point, respectively, and $\omega_{\rm a}$ is the acentric factor. Here $P_{\rm c}=2.204\times 10^{7}\,\mathrm{Pa}$, $T_{\rm c}=647.3\,\mathrm{K}$, and $\omega_{\rm a}=0.344$ for H$_{2}$O \citep[e.g.,][]{Chase1998}.

	The surface temperature is varied from 200 K to 4000 K during the course of the radiative transfer calculations. The temperature profile in the lower part of the atmosphere is supposed to follow an adiabatic lapse rate, and that in the upper part is given by the skin temperature, which is the asymptotic temperature at high altitudes of an upper atmosphere that is optically thin in the thermal-IR and transparent to shortwave radiation. The lapse rate is assumed to be a dry adiabat for atmospheric layers in which the temperature is above the critical temperature of H$_{2}$O or the relative humidity is less than 1. The dry adiabatic lapse rate is given by
	\begin{equation}
		\left(\frac{\partial T}{\partial P}\right)_{\rm dry}=\frac{\rho_{\rm v}T}{\rho_{\rm v}C_{p,\mathrm{v}}+\rho_{\rm n}C_{p,\mathrm{n}}}\left(\frac{\partial (1/\rho_{\rm v})}{\partial T}\right)_{P},
	\end{equation}
	where $\rho_{\rm v}$ and $\rho_{\rm n}$ are the densities of the condensible gas and non-condensible gas, respectively, $C_{p,\mathrm{v}}$ and $C_{p,\mathrm{n}}$ are the constant-pressure specific heat of the condensible gas and non-condensible gas, respectively \citep{Kasting1988, Marcq2012, Marcq2017}. For the layers where the relative humidity is unity, the lapse rate is given by a pseudo-moist adiabat as given by
	\begin{equation}
		\left(\frac{\partial T}{\partial P}\right)_{\rm moist}=\left[\frac{dP_{\rm sat}}{dT}+\frac{\rho_{\rm n}R}{M_{\rm n}}\left(1+\frac{d \mathrm{ln}\rho_{\rm v}}{d \mathrm{ln}T}-\frac{d \mathrm{ln}(\rho_{\rm v}/\rho_{\rm n})}{d \mathrm{ln}T}\right)\right]^{-1},
	\end{equation}
	where $P_{\rm sat}$ is the saturation vapor pressure of H$_{2}$O and $M_{\rm n}$ is the molecular weight of the non-condensible gas \citep{Kasting1988, Marcq2012, Wordsworth2013, Marcq2017}. Here,
	\begin{equation}
		\frac{d \mathrm{ln}(\rho_{\rm v}/\rho_{\rm n})}{d \mathrm{ln}T}=\frac{1}{(\rho_{\rm v}/\rho_{\rm n})(s_{\rm v}-s_{\rm c})+R/M_{\rm n}}\left[\frac{R}{M_{\rm n}}\frac{d \mathrm{ln}\rho_{\rm v}}{d \mathrm{ln}T}-C_{v,\mathrm{n}}-\frac{\rho_{\rm v}}{\rho_{\rm n}}\left(\frac{d s_{\rm v}}{d \mathrm{ln}T}\right)\right],
	\end{equation}
	where $s_{\rm v}$ and $s_{\rm c}$ are the specific entropies of the vapor phase and the condensed phase, respectively \citep{Kasting1988}, and $C_{v,\mathrm{n}}$ is the specific heat capacity at constant volume of the non-condensible gas. When the temperature reaches the skin temperature, the temperature above is assumed to be constant at the skin temperature as given by
	\begin{equation}
		T_{\rm skin}=\left(\frac{F_{\rm OLR}}{2\sigma}\right)^{1/4},
	\end{equation}
	where $\sigma$ is the Stefan-Boltzmann constant and $F_{\rm OLR}$ is the outgoing longwave radiation flux. The method to obtain $F_{\rm OLR}$ is described in the next subsection. We conducted iterative procedures to calculate $T_{\rm skin}$ and $F_{\rm OLR}$ self-consistently. 
	
	We refer to the temperature-dependent specific heat, entropy, and latent heat data provided in the NIST Chemistry WebBook \citep{Chase1998}. The saturation vapor pressure of H$_{2}$O is given by Osborne-Mayers and Washburn formula in the same manner as \citet{Abe1988}. At $273.15\,\mathrm{K} \leq T \leq 647.26\,\mathrm{K}$, 
	\begin{equation}
		\mathrm{ln}P_{\rm sat}(T)=\alpha - \frac{\beta}{t}+\frac{\gamma x}{t}[\mathrm{exp}(\delta x^{2})-1]-\epsilon \mathrm{exp}(-\eta y),
	\end{equation}
	where $t=T+0.01$, $x=t^{2}-293700$, $y=(647.26-T)^{5/4}$, $\alpha=24.021415$, $\beta=4616.9134$, $\gamma = 3.1934553\times 10^{-4}$, $\delta=2.7550431\times 10^{-11}$, $\epsilon=1.0246503\times 10^{-2}$, and $\eta=1.3158813\times 10^{-2}$ \citep{Eisenberg1969, Abe1988}. At $T \leq 273.15\,\mathrm{K}$,
	\begin{equation}
		\mathrm{ln}P_{\rm sat}(T)=\alpha'-\frac{\beta'}{t}+\gamma'\mathrm{log}t-\delta't+\epsilon't^{2},
	\end{equation}
	where $t=T-0.05$, $\alpha'=-10.666189$, $\beta'=5631.1206$, $\gamma' =18.953038$, $\delta'=3.8614490\times 10^{-2}$, and $\epsilon'=2.7749374\times 10^{-5}$ \citep{Abe1988}. 

\section{Details of the radiative transfer calculations}
	We solve the radiative transfer equation in a plane-parallel atmosphere on the two-stream approximation \citep{Toon1989} for the planetary and stellar radiation separately:
	\begin{equation}
		\frac{\partial F_{\nu}^{+}}{\partial \tau_{\nu}}=\gamma_{1}F_{\nu}^{+}-\gamma_{2}F_{\nu}^{-}-S_{\nu}^{+},
	\end{equation}
	\begin{equation}
		\frac{\partial F_{\nu}^{-}}{\partial \tau_{\nu}}=\gamma_{2}F_{\nu}^{+}-\gamma_{1}F_{\nu}^{-}+S_{\nu}^{-},
	\end{equation}
	where $\nu$ is the wavenumber, $\tau_{\nu}$ is the optical depth at $\nu$, $F_{\nu}^{+}$ and $F_{\nu}^{-}$ are the upward and downward radiative fluxes at $\nu$, and $\gamma_{i}$ is the coefficient which depends upon the particular form of the two-stream equation, respectively. For the planetary radiation,
	\begin{equation}
		S_{\nu}^{+}=S_{\nu}^{-}=2\pi (1-\omega_{0})B_{\nu}(T),
	\end{equation}
	where $\omega_{0}$ is the single scattering albedo and $B_{\nu}(T)$ is the Planck function. For the solar flux,
	\begin{equation}
		S_{\nu}^{+}=\gamma_{3}F^{\rm s}_{\nu}\omega_{0}\mathrm{exp}\left(-\frac{\tau_{\nu}}{\mu_{0}}\right),
	\end{equation}
	\begin{equation}
		S_{\nu}^{-}=(1-\gamma_{3})F^{\rm s}_{\nu}\omega_{0}\mathrm{exp}\left(-\frac{\tau_{\nu}}{\mu_{0}}\right),
	\end{equation}
	where $F^{\rm s}_{\nu}$ is the incoming solar flux at $\nu$, $\tau_{\nu}$ is the optical depth from the top of the atmosphere at $\nu$, and $\mu_{0}$ is the cosine of the mean stellar zenith angle ($57.3^{\circ}$) \citep[e.g.,][]{Hu2012}. Following the recommendation by \citet{Toon1989}, we use the hemispheric mean method for planetary radiation and the quadrature method for solar radiation. Here $\gamma_{1}=2-\omega_{0}$ and $\gamma_{2}=\omega_{0}$ in the hemispheric mean method, and $\gamma_{1}=\sqrt{3}(2-\omega_{0})/2$, $\gamma_{2}=\sqrt{3}\omega_{0}/2$, and $\gamma_{3}=1/2$ in the quadrature method \citep{Toon1989}.

	As radiative absorption sources, we consider line and continuum absorption of H$_{2}$O and collision-induced absorption of H$_{2}$-H$_{2}$. We use the H$_{2}$O line data in HITEMP 2010 \citep{Rothman2010}. The line profile of H$_{2}$O is assumed to follow a Voigt profile, truncated at 25 cm$^{-1}$ from the line center, to combine with continuum absorption represented by the MT\_CKD 4.2 model \citep{Mlawer2023} without double-counting. Here we use EXOCROSS \citep{Yurchenko2018} for the line profile calculations. We use the H$_{2}$-H$_{2}$ collision-induced absorption data provided by \citet{Borysow2001} and \citet{Borysow2002}. We also consider Rayleigh scattering by H$_{2}$O and H$_{2}$. The scattering cross-section is given by
	\begin{equation}
		\sigma^{\rm s}_{i}(\lambda)=\frac{128\pi^{5}(\alpha_{i}^{\rm p})^{2}}{3\lambda^{4}},
	\end{equation}
	where $\lambda$ is the wavelength, and $\sigma^{\rm s}_{i}$ and $\alpha^{\rm p}_{i}$ are the scattering cross-section and the polarizability constant of species $i$, respectively \citep[e.g.,][]{Pierrehumbert2010}. Here $\alpha^{\rm p}_{\rm H_{2}O}=1.50$ \AA$^3$ and $\alpha^{\rm p}_{\rm H_{2}}=0.787$ \AA$^3$, respectively \citep{Johnson1998}. Considering the absorption and scattering by the atmosphere, the optical depth $\tau_{\nu}$ is calculated by
	\begin{equation}
		\frac{d\tau_{\nu}}{dP}=\frac{1}{\rho g}\left[ \sum_{i} (\sigma^{\rm a}_{i}(\nu)+\sigma^{\rm s}_{i}(\nu))n_{i}+k_{\rm H_{2}-H_{2}}(\nu)n_{\rm H_{2}}^{2}\right],
	\end{equation}
	where $\rho$ is the atmospheric mass density, $g$ is the gravitational acceleration, $n_{i}$ is the number density of species $i$, $\sigma^{\rm a}_{i}(\nu)$ and $\sigma^{\rm s}_{i}(\nu)$ are the absorption and scattering cross-sections at $\nu$ of species $i$, and $k_{\rm H_{2}-H_{2}}(\nu)$ is the absorption coefficient of the H$_{2}$-H$_{2}$ collision-induced absorption at $\nu$, respectively.
	
	By calculating the scattered upward flux at the top of the atmosphere, we estimate the planetary albedo defined as follows:
	\begin{equation}
		A_{\rm p}=\frac{F^{+}_{\rm s,top}}{\mu_{0}S_{\odot}},
	\end{equation}
	where $A_{\rm p}$ is the planetary albedo, $F^{+}_{\rm s,top}$ is the scattered upward flux at the top of the atmosphere, and $S_{\odot}$ is the solar constant. Here we adopt the solar constant 0.7 times the present-day value considering the faint young Sun \citep{Gough1981}. The surface albedo is assumed to be 0.2 \citep[e.g.,][]{Kasting1988}. The net absorbed solar flux is given by
	\begin{equation}
		F_{\rm Sun}=\frac{1}{4}(1-A_{\rm p})S_{\odot}.
	\end{equation}
		
	We evaluate the integral of the radiative flux over a wavenumber range from 1 cm$^{-1}$ to 30000 cm$^{-1}$ with a resolution of 1 cm$^{-1}$ by applying a line-by-line method following \citet{Schaefer2016} and \citet{Wordsworth2017}. We tabulate the absorption cross-section on temperature and pressure grids. Here the temperature grids are [100, 150, 200, 250, 300, 350, 400, 450, 600, 800, 1000, 1250, 1500, 1750, 2000, 2250, 2500, 2750, 3000, 3250, 3500, 3750, 4000] K, and the pressure grids are [$10^{-5}$, $10^{-4}$, $5\times 10^{-4}$, $10^{-3}$, $5\times 10^{-3}$, $10^{-2}$, $5\times 10^{-2}$, $10^{-1}$, $5\times 10^{-1}$, 1, 5, 10, 50, $10^{2}$, $5\times 10^{2}$, $10^{3}$, $5\times 10^{3}$] bar. 

\section{Details of the interior structure and thermal evolution}
	We consider a spherically symmetric hydrostatic mantle where the lower boundary is the core-mantle boundary with a depth of 2900 km and the upper boundary is the planetary surface. The initial depth of the magma ocean is treated as a free parameter, corresponding to a basal pressure ranging from 1 to 100 GPa. The magma ocean is convecting vigorously, so the temperature distribution is supposed to be adiabatic following \citet{Lebrun2013}. As it cools, the adiabatic temperature profile intersects with the liquidus and solidus temperature profiles from the bottom. We use the solidus and liquidus curves of KLB-1 peridotite obtained from experimental data \citep{Hirschmann2000, Herzberg2000, Fiquet2010, Zhang1994}, whose fitting formula is derived by \citet{Nikolaou2019}. The melt fraction at any depth is given by
	\begin{equation}
		\phi = \frac{T-T_{\rm sol}}{T_{\rm liq}-T_{\rm sol}},
	\end{equation}
	where $\phi$ is the melt fraction, $T$ is the temperature of the mantle at a given depth, $T_{\rm sol}$ is the solidus temperature, and $T_{\rm liq}$ is the liquidus temperature, respectively \citep[e.g.,][]{Abe1997, Solomatov2007, Lebrun2013, Nikolaou2019}. The adiabatic temperature gradient is given by
	\begin{equation}
		\left(\frac{\partial T}{\partial P}\right)_{s} = \begin{cases}
			\frac{\alpha_{T} T}{\rho_{\rm melt} C_{p,\mathrm{MO}}},\hspace{10pt}\mathrm{for}\,\phi=1 \\
			\frac{\alpha_{T} T}{\rho_{\rm solid} C_{p,\mathrm{mantle}}},\hspace{10pt}\mathrm{for}\,\phi=0 \\
			\frac{\alpha_{T} T/\rho_{\rm MO}-\Delta H (\partial \phi/\partial P)_{T}}{C_{p,\mathrm{MO}}-\Delta H (\partial \phi/\partial T)_{P}},\hspace{10pt}\mathrm{for}\,0<\phi<1 \\
  		\end{cases}
	\end{equation}
	where $\rho_{\rm melt}$ is the melt density ($=4.0\times 10^{3}\,\mathrm{kg/m^{3}}$), $\rho_{\rm solid}$ is the solid density ($=4.2\times 10^{3}\,\mathrm{kg/m^{3}}$), $\rho_{\rm MO}$ is the mean density of the melt and solid in the magma ocean, $C_{p,\mathrm{MO}}$ and $C_{p,\mathrm{mantle}}$ are the specific heat of the magma ocean and mantle ($=1.0\times 10^{3}\,\mathrm{J/kg/K}$), respectively, $\Delta H$ is the difference of specific enthalpy between the melt and solid phases ($=4\times 10^{5}\,\mathrm{J/kg}$), and $\alpha_{T}$ is the thermal expansivity given by
	\begin{equation}
		\alpha_{T}=\alpha_{T,0}\left(\frac{PK'}{K_{0}}+1\right)^{-(m-1+K')/K'},
	\end{equation}
	where $\alpha_{T,0}=3\times 10^{-5}\,\mathrm{K^{-1}}$, $K_{0}=200\,\mathrm{GPa}$, $K'=4$, and $m=0$, respectively \citep{Abe1997, Lebrun2013}.
	
	The coupled thermal evolution is obtained by the energy conservation equation:
	\begin{equation}
		\int_{R_{\rm CMB}}^{R_{\rm p}}4\pi r^{2} \rho_{\rm int}(r) \left[C_{p,\mathrm{int}}(r)+\Delta H \left(\frac{\partial \phi}{\partial T}\right)_{P}\right]dr \frac{dT_{\rm s}}{dt}=4\pi R_{\rm p}^{2}(F_{\rm Sun}-F_{\rm OLR}),
	\end{equation}
	where $r$ is the radial distance from the planetary center, $R_{\rm CMB}$ is the radial distance of the core-mantle boundary, $R_{\rm p}$ is the planetary radius, $T_{\rm s}$ is the surface temperature, $F_{\rm OLR}$ is the outgoing planetary radiation flux, and $\rho_{\rm int}(r)$ and $C_{p,\mathrm{int}}(r)$ are the local density and specific heat capacity of the silicate interior at radial distance $r$, respectively \citep{Lebrun2013}. Here the cooling from the magma ocean is supposed to be along with the net energy balance of the incoming and outgoing radiation at the top of the atmosphere \citep{Hamano2013, Katyal2019}.



\end{document}